\documentclass[aps,pre,twocolumn,bibnotes,groupedaddress]{revtex4-1}		
\usepackage{amsmath,amsthm,amsfonts,amssymb,bm}
\usepackage{graphicx,subfigure}
\usepackage[dvipsnames]{xcolor}
\usepackage[]{natbib}
\usepackage[colorlinks=true,citecolor=blue]{hyperref}
\usepackage{epsfig}
\usepackage{epstopdf}

\numberwithin{equation}{section}

\newcommand{\vecv}[1]{\bm{{#1}}}
\newcommand{\tens}[1]{\bm{{#1}}}

\newcommand{\partials}[2]{\partial_{#2}{#1}}

\newcommand{\be}{\begin{equation}}
\newcommand{\ee}{\end{equation}}
\newcommand{\bea}{\begin{eqnarray}}
\newcommand{\eea}{\end{eqnarray}}
\newcommand{\EC}{\partials{F}{\epsilon}_{\rm elastic}}

\newcommand{\edotbar} { \bar{\dot{\epsilon}}}
\newcommand{\ebar} { \bar{\epsilon}}
\newcommand{\edot} { \dot{\epsilon}}
\newcommand{\eddot} { \dot{\dot{\epsilon}}}

\newcommand{\sigmae} { \sigma_{\rm E}}
\newcommand{\sigmaei} { \sigma_{\rm E0}}
\newcommand{\sigmaeb} { \sigma_{\rm E0}}
\newcommand{\force} { F}
\newcommand{\ftrans}{\tilde{F}}
\newcommand{\considere} {Consid\`ere }

\newcommand{\Fi}{F_c}

\newcommand{\taur}{\tau_R}
\newcommand{\taud}{\tau_d}
\newcommand{\taus}{\tau_s}
\newcommand{\taub}{\tau_b}

\newcommand{\conf}{\tens{W}} 
\newcommand{\total}{\tens{T}}
\newcommand{\visc}{\tens{\Sigma}}

\begin{document}
\title{Criteria for extensional necking instability in complex fluids
  and soft solids. \newline Part II: imposed tensile stress and force protocols.}

\author{D. M. Hoyle} 
\email{d.m.hoyle@durham.ac.uk}
\homepage{http://community.dur.ac.uk/d.m.hoyle/ }
\affiliation{Department of Physics, University of Durham, Science Laboratories, South Road,
  Durham, DH1 3LE, United Kingdom}

\author{S. M. Fielding} 
\email{suzanne.fielding@durham.ac.uk}
\homepage{http://community.dur.ac.uk/suzanne.fielding/}
\affiliation{Department of Physics, University of Durham, Science
  Laboratories, South Road, Durham, DH1 3LE, United Kingdom}

\begin{abstract}

  We study the necking of a filament of complex fluid or soft solid
  subject to uniaxial tensile stretching, separately under conditions
  of constant imposed tensile stress and constant imposed tensile
  force, by means of linear stability analysis and nonlinear
  simulations at the level of a slender filament approximation. We
  demonstrate necking to be a flow instability that arises as an
  unavoidable consequence of the viscoelastic constitutive behaviour
  of essentially any material (with a possible rare exception).  We
  derive criteria for the onset of necking that can be reported in
  terms of characteristic signatures in the shapes of the
  experimentally measured material functions, and that should
  therefore apply universally to all viscoelastic materials.  To
  confirm their generality, we show them to hold numerically in six
  constitutive models: the Oldroyd B, Giesekus, non-stretch
  Rolie-Poly, finite-stretch Rolie-Poly and Pom-pom models, and a
  simplified toy model of coil-stretch hysteresis, which has a
  non-monotonic underlying extensional constitutive curve.  Under
  conditions of constant imposed tensile stress, we find two distinct
  dynamical regimes as a function of the time since the inception of
  the flow.  In the first regime the strain rate quickly attains a
  value prescribed by the fluid's underlying stationary homogeneous
  extensional constitutive curve, at the given imposed stress. During
  this first regime, no appreciable (or only minimal) necking arises.
  A second regime then ensues in which the initially homogeneous flow
  destabilises to form a neck.  This necking instability can occur via
  two distinct possible modes.  The first mode is relatively gentle and
  arises in any regime where the slope of the extensional constitutive
  curve is positive. It has a rate of necking per accumulated strain
  unit set by the inverse of the slope of the constitutive curve on a
  log-log plot. The second mode sets in when a carefully defined
  `elastic derivative' of the tensile force first slopes down as a
  function of the time since the inception of the flow.  We discuss
  the way in which these modes of instability manifest themselves in
  entangled polymeric fluids, demonstrating four distinct regimes of
  necking behaviour as a function of imposed stress. Under conditions
  of constant imposed tensile force, typically the flow sweeps up the
  underlying constitutive curve of the fluid in question, again with
  instability to necking in any regime where that curve is positively
  sloping.

\end{abstract}

\date{\today}
\maketitle

\section{Introduction}
\label{sec:intro}

In developing a constitutive theory for the rheology of a complex
fluid or soft solid, key aims are to predict the material's stress
response as a function of the applied strain history (or vice versa),
in both shear and extension. In comparison with shear, extensional
flows typically subject the underlying fluid microstructure (polymer
chains, wormlike micelles, {\it etc.}) to more severe reorganisation.
As a consequence, many nonlinear flow features manifest themselves
only in extension. In this way, extensional flows prove crucial in
discriminating between alternative possible constitutive theories.

To characterise a material's extensional rheology experimentally, a
common procedure consists of progressively stretching out in length an
initially undeformed cylindrical sample. For a review,
see~\cite{McKinley2002b}. In stretching at constant imposed Hencky
strain rate $\edot$, the most commonly reported rheological response
function is the tensile stress $\sigmae^+(t)$ as a function of the
time $t$ since the inception of the flow.  If this can be measured to
steady state, the steady tensile stress $\sigmae^+(t\to\infty)$
plotted as a function of applied strain rate $\edot$, obtained in a
series of filament stretching runs performed at different strain
rates, then gives the material's extensional constitutive curve, or
flow curve, $\sigmae(\edot)$.  Another common protocol consists of
stretching a filament under conditions of a constant imposed tensile
stress
$\sigmae$~\cite{Munstedt1975,Munstedt2013,Munstedt2014,Kurzbeck1999a,Laun1976,Laun1978,Munstedt2005,Munstedt1979,Wolff2011}.
This typically allows a fluid to attain a stationary flow prescribed
by its constitutive curve more readily than stretching at a constant
strain rate, making it a preferable protocol for measuring that
constitutive curve experimentally~\cite{Alvarez2013}.  Stretching at
constant tensile force $\force$ is also commonly
performed~\cite{Wagner2002,Szabo2012,Matta1990,Raible1982,Wagner2012}.

In these filament stretching procedures, an important aspiration is to
ensure that the flow remains as homogeneous as possible, in some part
of the sample at least, to allow the measurement of the steady state
homogeneous constitutive properties just described.  Almost
ubiquitously observed during filament stretching, however, is the
onset of heterogeneous deformation along the length of the filament.
Typically the central region, furthest from the sample ends, develops
a higher strain rate than the globally averaged one and thins more
quickly than the sample as a whole, hindering attempts to characterise
the fluid's homogeneous constitutive properties.

In a typical filament stretching rheometer, the sources of this flow
heterogeneity are (at least) twofold.  The first is essentially
imposed on the fluid externally by the flow geometry, by the no-slip
boundary condition that pertains where the fluid makes contact with
the rheometer plates. This prevents those parts of the sample that are
nearest the plates from being properly stretched.  In consequence, the
central regions of the filament, furthest from the plates, thin more
quickly than the sample as a whole.

The second source of heterogeneity, in contrast, is intrinsic to the
constitutive behaviour of the fluid. It takes the form of a
hydrodynamic necking instability, which can be described in simple
terms as follows \cite{Fielding2011,strainPaper}.  Consider a filament that is initially stretching in a
purely uniform way. Then imagine a local upward fluctuation in strain
rate at some location along the filament.  Conservation of mass means
that the filament must then thin faster in that location. To maintain
force balance along the filament a counterbalancing larger stress must
then develop in this location. To achieve this larger stress, the
fluid must flow faster at that location.  This enhances the original
fluctuation, giving a positive feedback loop and a runaway instability
in which that part of the filament thins more quickly than the
filament as a whole, forming a neck.

In practice, of course, these two sources of heterogeneity interact
with each other.  In particular, the extrinsic geometrical
heterogeneity imposed by the no-slip condition at the end-plates
provides an initial `seed' that is then picked up and amplified by the
intrinsic hydrodynamic necking instability just described.  The
necking instability described eventually causes the filament to fail
altogether, aborting the experimental run.  It has been seen in linear
polymers \cite{Barroso2005a}, branched polymers
\cite{Liu2013,Burghelea2011a}, associative polymers
\cite{Tripathi2006}, wormlike micelles \cite{Bhardwaj2007}, bubble
rafts \cite{Arciniaga2011}, and dense colloidal suspensions
\cite{Smith2010}. It arises in all common stretching protocols,
including at constant tensile stress \cite{Andrade2011}, constant
tensile force~\cite{Wagner2012}, constant imposed
Hencky strain rate \cite{Burghelea2011a,Malkin2014}, and following a
finite Hencky strain ramp~\cite{Wang2007}.

In a separate manuscript, we studied theoretically the instability to 
necking of a filament of complex fluid or soft solid subject to stretching 
under conditions of constant imposed Hencky strain rate \cite{strainPaper}. 
We considered several different constitutive models, with the aim of 
modelling several different classes of soft material, including: polymer 
solutions, polymer melts of both linear and branched chain architectures, 
worm-like micelles and soft glassy materials. Universally across these 
materials, we found there to be two distinct possible modes of necking instability. 
The first mode, which has a modest associated growth rate, first becomes 
unstable in a filament stretching experiment when the extensional stress 
signal first attains a negative curvature as a function of time since 
the inception of the flow (or, equivalently, as a function of the accumulated 
Hencky strain). The second mode, which leads to much more rapid necking, 
first becomes unstable when a carefully defined `elastic' derivative of 
the tensile force with respect to Hencky strain first becomes negative.  
In the limit of infinite imposed strain rate, we showed that this second 
mode coincides with the well known \considere criterion for necking in 
solids \cite{Considere1885}, which predicts instability when the conventionally defined 
derivative of the tensile force with respect to Hencky strain first 
becomes negative. Importantly, however, we also showed that this 
original \considere criterion fails to correctly predict the 
onset of necking at finite imposed strain rates.

In this work, we perform the counterpart
analysis for filament stretching at a constant imposed tensile stress
and (separately) a constant imposed tensile force.  (Although this
manuscript is intended to be self-contained in its own right, it would
most effectively be read alongside its counterpart
in~\cite{strainPaper}. Inevitably, in some places, particularly the
introductory sections, the discussion in the present manuscript
mirrors that of the controlled strain one.)

Our aims are fourfold. First, we seek to demonstrate that necking is a
flow instability that arises as an inevitable consequence of the
constitutive behaviour of essentially any complex fluid or soft solid.
However we also identify a possible rare exception in which a fluid
may be stable against necking, and discuss it carefully.

Second, we shall derive criteria for the onset of necking that are
universal to all complex fluids and soft solids, and are reportable
simply in terms of characteristic signatures in the shapes of the
experimentally measured rheological material response functions. We
shall first derive these criteria analytically by means of linear
stability calculations performed in a constitutive model of a highly
simplified and generalised form.  We shall then confirm their
generality by performing numerical simulations in six concrete choices
of constitutive model: the Oldroyd B, Giesekus~\cite{Larson1988},
non-stretch Rolie-Poly~\cite{Likhtman2003}, finite-stretch
Rolie-Poly~\cite{Likhtman2003} and Pom-Pom
models~\cite{Blackwell2000,McLeish1998}, and a simplified model of
coil-stretch hysteresis~\cite{Somani2010,Schroeder2003,DeGennes1974}, which has a non-monotonic
underlying extensional constitutive curve.

Third, we seek to elucidate the way in which these criteria manifest
themselves in entangled polymeric fluids. In particular, in filament
stretching experiments performed at constant imposed tensile stress,
we predict four distinct regimes of necking as a function of the
imposed stress.
Fourth, we shall show that the \considere criterion for necking in
solids entirely fails to predict necking in complex fluids, and must
instead be replaced by the criteria offered here.

Throughout we shall restrict ourselves to the case of a highly a
viscoelastic filament of sufficiently large (initial) radius that bulk
viscoelastic stresses dominate surface effects in determining the
onset of necking. Accordingly, we set the surface tension to zero in
our calculations. We therefore do not address capillary breakup as
studied in CaBeR rheometers
\cite{Clasen2006,Tembely2012,Szabo2012,Vadillo2012,Webster2008,McIlroy2014,Bhat2008}.

All our calculations assume a slender filament approximation in which
the wavelengths of any heterogeneities along the filament's length are
long compared to the filament radius.  Our calculation therefore
cannot capture the final pinch-off of any neck (which would surely in
any case be affected by surface tension), nor can it address a more
dramatic fracture mode in which the filament sharply rips across its
cross section.

The manuscript is structured as follows. We start in
Sec.~\ref{sec:models} with a discussion of our theoretical framework.
In particular we introduce the constitutive models and flow geometry
under consideration, as well as the slender filament approximation and
an exact transformation to the frame that co-extends and co-thins with
the filament as it is stretched out. We also outline our linear
stability analysis, and the numerical method of our nonlinear slender
filament simulations.  In Secs.~\ref{sec:stress} and~\ref{sec:force}
we present results for the protocols of constant imposed tensile
stress and constant imposed tensile force respectively. In
Sec.~\ref{sec:conclusions} we offer a summary and perspectives for
future work.

\section{Theoretical framework}
\label{sec:models}

\subsection{Mass balance and force balance}
\label{sec:balance}

We assume the total stress $\total(\tens{r},t)$ at time $t$ in a fluid
element at position vector $\tens{r}$ to comprise the sum of a
viscoelastic contribution $\visc(\tens{r},t)$ from the internal fluid
microstructure (polymer chains, wormlike micelles, {\it etc.}), a
Newtonian contribution of viscosity $\eta$ arising from the solvent or
other fast degrees of freedom, and an isotropic contribution with a
pressure $p(\tens{r},t)$:
\be
\total = \visc + 2 \eta \tens{D} - p\tens{I}.
\label{eqn:total_stress_tensor}
\ee
The symmetric strain rate tensor $\tens{D} = \frac{1}{2}(\tens{K} +
\tens{K}^T)$ where $K_{\alpha\beta} =
\partial_{\beta}v_{\alpha}$ and $\tens{v}(\tens{r},t)$ is the fluid
velocity field.  Throughout we shall work in the limit of zero
Reynolds number, assuming conditions of creeping flow in which force
balance requires:
\be
\vecv{\nabla}\cdot\,\total = 0.
\label{eqn:force_balance}
\ee
We also assume incompressible flow, with the pressure field
$p(\tens{r},t)$ determined by enforcing 
\be
\label{eqn:incomp}
\vecv{\nabla}\cdot\vecv{v} = 0.
\ee

\subsection{Constitutive models}
\label{sec:constitutive}

The dynamics of the viscoelastic stress $\visc$ contributed by the
internal fluid microstructure (polymer chains, {\it etc.}) is
specified by a constitutive model for the fluid in question. In this
work we shall consider several different constitutive models, most of
which are fully tensorial and widely used throughout the rheological
literature.  We also invoke a simplified scalar constitutive model to
allow analytical progress in deriving criteria for the onset of
necking. We shall then check that the predictions obtained in that
model also hold numerically in the tensorial models.  We summarise the
models now in turn.

\subsubsection{Oldroyd B model}
\label{app:OldB}

The Oldroyd B model provides a phenomenological description of the
rheology of dilute polymer solutions. It represents each polymer chain
as a dumbbell comprising two beads connected by a Hookean spring.  A
conformation tensor $\conf=\langle \vecv{R}\vecv{R}\rangle$ is defined
as the ensemble average $\langle \rangle$ of the outer dyad of the
dumbbell end-to-end vector $\vecv{R}$, which is taken to have unit
length in the absence of flow. The viscoelastic stress is assumed to
depend on the conformation tensor according to
\be
\label{eqn:Hooke}
\visc = G\left( \conf - \mathbf{I} \right),
\ee
with a constant modulus $G$.  The dynamics of the conformation tensor
 obeys
\be
\overset{\nabla}{\mathbf{W}} = -\frac{1}{\tau} \left( \mathbf{W} - \mathbf{I} \right),
\label{eqn:Maxwell}
\ee
with a characteristic relaxation time $\tau$. The upper convected
derivative
\be
\overset{\nabla}{\mathbf{W}} = \frac{D\conf}{Dt} - \conf\cdot\mathbf{K} - \mathbf{K}^T\cdot\conf,
\ee
with a velocity gradient tensor
$\mathbf{K}_{\alpha\beta}=\partial_\alpha v_\beta$. This in turn
contains the Lagrangian derivative
\be
\frac{D\conf}{Dt} = \frac{\partial \conf}{\partial t} + \mathbf{v}\cdot\nabla\conf.
\ee

For an imposed uniaxial extensional flow along the Cartesian $z-$axis
we have
\be
\tens{K}=\edot\begin{pmatrix} -\frac{1}{2} & 0 & 0 \\
                           0 & -\frac{1}{2} & 0 \\
                           0 & 0 & 1 \\
	\end{pmatrix}.
\ee
For any sustained imposed strain rate $\edot> 1/2\tau$ the Oldroyd B
model predicts a dynamical catastrophe in which the dumbbells stretch out
indefinitely and the extensional stress diverges. Its extensional
constitutive curve, which gives the relationship between the tensile
stress $\sigmae=G(W_{zz}-W_{xx})+3\eta\edot$ and strain rate $\edot$
in a stationary flow, is therefore undefined for $\edot>1/2\tau$. See
Fig.~\ref{fig:constitutive}a).

\subsubsection{Giesekus model}
\label{app:Giesekus}

The Giesekus model describes more concentrated polymer solutions, by
generalising the Oldroyd B model to postulate an anisotropic drag
such that the relaxation time of a dumbbell is altered when the
surrounding dumbbells are oriented~\cite{Larson1988}.  The dependence
of the stress on the conformation tensor remains as in
Eqn.~\ref{eqn:Hooke}, but the conformation tensor now obeys modified
dynamics:
\be 
\overset{\nabla}{\conf} = - \frac{1}{\tau} \left( \conf
  - \mathbf{I} \right) - \frac{\alpha}{\tau} \left( \conf -
  \mathbf{I} \right)^2.
\ee

The parameter $\alpha$ lies in the range $0 \le \alpha\le 1$, with
Oldroyd B dynamics recovered when $\alpha= 0$. For $\alpha > 0$ the
extensional catastrophe of Oldroyd B is averted by the anisotropic
drag: the Giesekus model has a well defined constitutive curve at all
extension rates. See Fig.~\ref{fig:constitutive}b).

\subsubsection{Rolie-Poly model of entangled linear polymers} 
\label{app:RP}

As a description of more concentrated solutions or melts of entangled
linear polymers, we used the Rolie-Poly model~\cite{Likhtman2003}.
This starts from microscopic considerations based on the tube theory
of polymer dynamics~\cite{Doi1986}, whereby any polymer chain is
assumed to be dynamically restricted by a tube of entanglements with
nearby chains. If then refreshes its configuration by a process of 1D
curvilinear diffusion (``reptation'') along the tube contour. Also
included are the additional dynamical processes of chain stretch
relaxation and convective constraint release
\cite{Marrucci1996,Ianniruberto2014,Ianniruberto2014a}, in which the
relaxation of the stretch of any chain acts also to relax entanglement
points with other chains, and so facilitate the relaxation of
orientation.  These processes are modelled via a differential
constitutive equation for the the dynamics of the conformation tensor
$\conf=\langle \vecv{R}\vecv{R}\rangle$, with $\vecv{R}$ the
end-to-end vector of a polymer chain:
\begin{widetext}
\be
\overset{\nabla}{\conf} = - \frac{1}{\tau_d} \left( \conf - \mathbf{I} \right) 
  - \frac{2}{\taus\left( 1 - f T/3 \right)}\left( 1 - \sqrt{\frac{3}{T}}\right)
\left[ \conf  + \beta\left(\frac{T}{3}\right)^{\delta}( \conf - \mathbf{I} ) \right],
\ee
\label{eqn:sRP}
\end{widetext}
where the trace $T=\sum_i W_{ii}$. In this equation, $\taud$ and
$\taus$ are the characteristic timescales of reptation and
chain-stretch relaxation respectively.  These are assumed to be in the
ratio
\be
\frac{\taud}{\taus}=3Z,
\label{eqn::3Z}
\ee
where $Z$ is the number of entanglements per chain.  The parameter
$\beta$ in Eqn.~\ref{eqn:sRP} sets the degree of convective constraint
release, while the factor $(1-f T/3)$ accounts for finite chain
extensibility. (For $f=0$ the model predicts an Oldroyd B-like stretch
catastrophe for a sustained strain rate $\edot>1/\taus$.)

For a highly entangled sample, with a large entanglement number $Z$,
the chain-stretch relaxes very quickly on the timescale of reptation,
$\taus\ll\taud$. For imposed flow rates $\edot\ll 1/\taus$ we can then
take the limit $\taus \to 0$ upfront and use the simpler,
non-stretching form of the model:
\be
  \overset{\nabla}{\conf} 
= -\frac{1}{\taud}(\conf-\mathbf{I})-\frac{2}{3}\mathbf{K}:\conf\,(\conf+\beta(\conf-\mathbf{I})).
\ee
For a value $\beta=0$ of the convective constraint release parameter,
this also recovers the reptation-reaction model of entangled wormlike
micelles~\cite{Cates1990}.

\begin{table*}
	\centering
	\begin{tabular*}{0.7\textwidth}{@{\extracolsep{\fill}} c | c  }
		\hline
		\hline
		Model & Parameters \\
		\hline
		Oldroyd-B & - \\
		Giesekus & $\alpha = 0.001$  \\
		Non-stretch Rolie-Poly & $\beta=0.0$ \\
		Finite-stretch Rolie-Poly & $\beta=0.0$, $\delta=-0.5$, $\tau_R = 0.00833$ and $f = 0.000625$ ($Z=40$)\\
		Pom-pom  & $\tau_s = 0.1$ and $q = 40$ \\
		Hysteresis  & $c=0.001$\\
		\hline
		\hline
	\end{tabular*}
	\caption{Parameter values used in our numerical studies.  The solvent viscosity is taken to obey $3\eta=0.01$ in all models.}
\label{tab::model_params}
\end{table*}

\subsubsection{Pom-pom model of entangled branched polymers}
\label{app:pom-pom}

As a description of the rheology of entangled long-chain branched
polymers, we use the Pom-pom model~\cite{Blackwell2000,McLeish1998}.
Each molecule is taken to comprise a linear backbone with an equal
number of arms $q$ attached to each end. The relaxation of the arms is
assumed to be fast compared to that of the backbone, acting only to
provide an additional drag on the backbone dynamics.

The viscoelastic stress is then taken to depend on a conformation
tensor $\conf$ that specifies the orientation of the backbone, and the
degree of backbone stretch $\lambda$:
\be
\mathbf{\Sigma} = 3G\lambda^2 \left( \mathbf{W} -  \frac{1}{3}\mathbf{I} \right).
\ee

The dynamics of the backbone orientation is modeled by writing
\begin{equation}
\mathbf{W} = \frac{\mathbf{A}}{\rm{tr}(\mathbf{A})},
\end{equation}
with the dynamics of $\mathbf{A}$ then assumed to obey the Maxwell
model, Eqn.~\ref{eqn:Maxwell}, with relaxation time $\taub$.  The
backbone stretch has dynamics
\be
\label{eqn:lambda}
\frac{D\lambda}{Dt} = \lambda \mathbf{K}:\mathbf{W} - \frac{1}{\tau_s}\left( \lambda - 1 \right) e^{\nu^*\left( \lambda - 1 \right) }\;\;\textrm{for}\;\;\;\lambda\le q,
\ee
where $\nu^* = 2/(q-1)$, subject to an initial condition $\lambda(0) =
1$. A hard upper cutoff is imposed at $\lambda=q$: the extent of
backbone stretch is assumed to be entropically bounded by the number
of arms attached to each end of the backbone.  This hard cutoff leads
to catastrophically fast necking in this version of the Pom-pom model,
both in the constant imposed Hencky strain rate protocol studied in
Ref.~\cite{strainPaper}, and in the constant imposed tensile stress protocol
considered below.

The timescales of the backbone orientation and stretch dynamics are
assumed to be in the ratio
\be
\frac{\taub}{\taus} = Z_b\phi_b.
\label{eqn::phibZ}
\ee
Here $Z_b$ is the number of entanglements along the backbone and
\be
\phi_b = \frac{Z_b}{Z_b + 2qZ_a},
\ee
which is the fraction of material in the backbone compared with that
in the molecule as a whole, with $Z_a$ the number of entanglements
along each arm.

\subsubsection{Generalised scalar constitutive model}
\label{app:Toy-Model}

So far, we have outlined the tensorial constitutive models to be
studied numerically in the rest of the paper. To allow analytical
progress in deriving criteria for the onset of necking we also
consider a simplified scalar model~\cite{strainPaper}, which assumes
the dominant component of microstructural deformation that develops in
a filament stretching experiment to be $W_{zz}$, where $z$ is the
coordinate along the length of the filament. We denote $W_{zz}=Z$ for
notational simplicity. (This should not be confused with our use of
$Z$ to denote entanglement number in the polymer models above.)

We then consider highly generalised constitutive dynamics for $Z$, following
\be
\label{eqn:toy}
\frac{DZ}{Dt}=\edot f(Z)-\frac{1}{\tau}g(Z),
\ee
with separate loading and relaxation terms characterised by the
functions $f$ and $g$ respectively. We intentionally write the model
in this highly generalised way, without specifying particular
functional forms for the loading and relaxation dynamics $f(Z)$ and
$g(Z)$. Our aim in so doing will be to derive criteria for the onset
of necking that are reportable simply in terms of characteristic
signatures in the shapes of the material's bulk rheological response
functions.

The tensile stress then comprises the usual sum of this viscoelastic
component and a Newtonian (solvent) contribution:
\be
\label{eqn:stressToy}
\sigmae=G Z + \eta\edot,
\ee
in which for simplicity, in this scalar model, we have absorbed a
factor $3$ into the Newtonian viscosity $\eta$.

\subsubsection{Toy scalar hysteresis model}

As a simple model of coil-stretch hysteresis~\cite{Somani2010,Schroeder2003,DeGennes1974}, which
has been observed in polymer solutions, and which is associated with a
non-monoonic constitutive curve, we use the scalar model introduced in
the previous section, with particular functional choices for $f$ and
$g$:
\be
f=3+2Z
\ee
and
\be
g=\frac{Z}{1+Z^{3/2}}+c Z^2.
\ee

\subsection{Units and parameter values.}
\label{sec:units}

We use units of length in which the initial length of the filament
$L(0)=1$, and of stress in which the viscoelastic modulus $G=1$.  In
any given model, units of time are adopted in which the viscoelastic
relaxation timescale is equal to unity. Accordingly for the Oldroyd B
and Giesekus models we set $\tau=1$, the Rolie-Poly model $\taud=1$,
and for Pom-pom model $\taub=1$.
Values for the other model parameters, in these units, are listed in
table~\ref{tab::model_params}.

\subsection{Initial conditions, flow geometry and protocol.}
\label{sec:geometry}

We consider a filament that at some initial time $t=0$ is in the form
of an undeformed uniform cylinder of length $L(0)$ in the direction
$z$ along the length of the cylinder, and cross sectional area $A(0)$
in the $xy$ plane.  The viscoelastic stresses in the material are
assumed to be well relaxed in this initial state, such that the
molecular conformation tensor $\conf(0)=\tens{I}$.  At time $t=0$ the
filament is then subject to the switch-on of either a tensile stress
or a tensile force, which is held constant thereafter.

As a result of this imposed load, the filament will progressively
stretch out in length according to some creep curve in the Hencky
strain signal $\ebar(t)$, with the filament length accordingly
increasing as $L(t)=L(0)\exp(\ebar(t))$.  The overbar signifies that
$\ebar$ is the nominal Hencky strain experienced by the sample as a
whole: {\it i.e.}, the Hencky strain averaged along the length of the
filament.  Once necking arises, the strain and strain rate will
locally vary along the filament's length $z$. For example the Hencky
strain rate $\edot=\edot(z,t)$, with a $z-$averaged $\edot$ equal to
$\edotbar$.

\subsection{Slender filament approximation}
\label{sec:slender}

We adopt a slender filament approximation \cite{Forest1990,
  Olagunju1999, Denn1975}, in which the characteristic wavelengths of
any variations in cross sectional area that develop along the
filament's length during necking are taken to be large compared to the
filament's radius. This allows us to average the flow variables over
the filament's cross section at any location $z$ along it.  The
relevant dynamical variables are then the cross sectional area
$A(z,t)$, the area-averaged fluid velocity in the $z$ direction
$V(z,t)$, the extension rate $\edot(z,t)=\partial_z V$, and any
relevant viscoelastic variables contained in the constitutive
equations discussed in the previous section.

Within this approximation, the mass balance condition
(\ref{eqn:incomp}) is written
\be
\partials{A}{t} + V\partials{A}{z}  = -\dot\varepsilon A, \label{eqn:1Dmass} 
\ee
and the force balance condition (\ref{eqn:force_balance})  
\be
0 = \partials{F}{z},
 \label{eqn:1Dmom} 
\ee
in which the tensile force
\be
\label{eqn:tforce}
F(t)=A(z,t)\sigmae(z,t),
\ee
and the total tensile stress 
\be
\label{eqn:tstress}
\sigmae = G\left(W_{zz} - W_{xx} \right) + 3\eta\dot\varepsilon.  \ee
The Lagrangian derivative on the left hand side of any constitutive
equation is now written:
\be
\frac{D}{Dt}=\frac{\partial}{\partial t} + V\frac{\partial}{\partial z}.
\ee

Without sacrificing generality we  set the initial cylinder area
$A(0)=a_0=1$.  Although this is in addition to having set the initial
cylinder length $L(0)=1$ (recall Sec.~\ref{sec:units} above) we
emphasize that we are not, in fact, restricting ourselves to scenarios
in which the initial area and length are constrained relative to each
other in any particular way.  Any information about their relative
values has simply been lost in making the slender filament
approximation.

\subsection{Transformation to co-extending frame}

As the filament stretches out under the imposed tensile stress (or
force) its length increases in time as $L(t)=L(0)\exp(\ebar(t))$, and
the length-averaged area decreases as $A(t)=A(0)\exp(-\ebar(t))$,
where $\ebar(t)$ is the nominal (length-averaged) Hencky strain. To
allow for this overall exponential change in the filament's shape as a
function of the accumulating strain, it is convenient to make a
coordinate transformation to the coextending, cothinning frame by
defining new variables of length $u$, velocity $v$ and area $a$ as
follows:
\begin{eqnarray}
		u &=& z\exp(-\ebar(t)),	\nonumber \\
		v(u,t) &=& V(z,t)\exp(-\ebar(t)),	\nonumber \\
		a(u,t) &=& A(z,t)\exp( \ebar(t)).
\end{eqnarray}
The differential operators  transform as
\begin{eqnarray}
\partial_z &\longrightarrow &  \exp(-\ebar(t))\partial_u,	\\
\partial_t &\longrightarrow &  \partial_t - \edotbar u \partial_u.
\end{eqnarray}
We then have transformed equations of mass balance  
\be
\label{eqn:massT}
\partials{a}{t} + (v-\edotbar u)\partials{a}{u} = -(\edot-\edotbar)a,
\ee
and force balance 
\be
\label{eqn:forceT}
 0 = \partials{\ftrans}{u},
\ee
where the transformed tensile force 
\be
\label{eqn:stressT}
\ftrans(t)=F(t)\exp(\ebar(t))=a(u,t)\sigmae(u,t).
\ee
The tensile stress $\sigmae$ is given as in~\ref{eqn:tstress} above.
The Lagrangian derivative on the left hand side of any viscoelastic
constitutive equation then becomes
\be
\label{eqn:confT}
\frac{D}{Dt} = \frac{\partial}{\partial t}  + (v-\edotbar u)\frac{\partial}{\partial u}.
\ee
The other (local) terms of the constitutive equations are unaffected
by the transformation.

\subsection{Linear stability analysis}
\label{sec:LSA}

We now discuss our linear stability analysis to determine the onset of
necking.  This starts by considering a homogeneous ``base state''
corresponding to a filament that remains a uniform cylinder as it is
stretched out, with the flow variables remaining uniform along it.  We
then add to this base state small amplitude perturbations describing
any slight initial heterogeneities along the filament's length, which
are the precursor of a neck. (We return below to discuss possible
sources for these heterogeneities, in particular in the no-slip
condition where the sample meets the rheometer endplates.)  Expanding
the governing equations to first order in the amplitude of these
perturbations then allows us to arrive at a set of linearised
equations for the dynamics of the perturbations. Our interest is then
in determining whether, and at what time during filament stretching,
the perturbations grow into a necked state, or whether they decay to
leave a uniform filament.

To allow analytical progress, we shall flesh out the details of the
procedure just described in the context of the scalar constitutive
model of Sec.~\ref{app:Toy-Model}. The entirely analogous (but more
cumbersome) calculation for the fully tensorial models is described
in detail in Sec.~IV of Ref.~\cite{strainPaper}. It is that calculation which
underpins our numerical results in Secs.~\ref{sec:stress}
and~\ref{sec:force} below.

Consider first, then, a uniform base state corresponding to a filament
that remains a perfect cylinder as it is stretched out, with all flow
variables homogeneous along it. In the laboratory frame this obeys the
homogeneous form of Eqns.~\ref{eqn:1Dmass} to~\ref{eqn:tstress} above,
together with the homogeneous form of the scalar constitutive
equation. Accordingly the condition of mass
balance gives
\be
\label{eqn:baseMass}
\dot{A}_0(t)=-\edot_0 A_0.
\ee
(In the cothinning frame, the base state area $a_0$ is obviously
constant in time by definition.)  The tensile force
\be
F_0(t)=A_0\sigma_0,
\ee
with tensile stress
\be
\sigma_0(t)=GZ_0+\eta\edot_0.
\ee
The viscoelastic variable evolves according to
\be
\label{eqn:baseZ}
\dot{Z}_0(t)=\edot_0f(Z_0)-\frac{1}{\tau}g(Z_0).
\ee
To distinguish this uniform base state from the heterogeneous
perturbations to it that we shall go on to consider, we have labelled
its flow variables with a subscript $0$.

To allow for the possibility of necking, we must now account for
spatial variations along the filament's length. Accordingly we return
to the spatially aware form of the model equations, expressed for
convenience in the co-thinning, co-extending frame:
Eqns.~\ref{eqn:massT} to~\ref{eqn:stressT}, together with the scalar
constitutive model, Eqn.~\ref{eqn:toy}. We collect these again here
for convenience.

The condition of mass balance gives
\be
\label{eqn:massT2}
\partials{a}{t} + (v-\edotbar u)\partials{a}{u} = -(\edot-\edotbar)a,
\ee
while force balance  gives
\be
\label{eqn:forceT2}
 0 = \partials{\ftrans}{u},
\ee
with the transformed tensile force 
\be
\label{eqn:stressT2}
\ftrans(t)=F(t)\exp(\ebar(t))=a(u,t)\sigmae(u,t).
\ee
The tensile stress is given by 
\be
\sigmae(u,t)=GZ(u,t)+\eta \edot(u,t),
\ee
and the viscoelastic variable evolves according to
\be
\label{eqn:visc2}
\frac{\partial Z}{\partial t}  + (v-\edotbar u)\frac{\partial Z}{\partial u}=\edot f(Z)-\frac{1}{\tau}g(Z).
\ee

We now add to the homogeneous base state small amplitude heterogeneous
perturbations, which are the precursor of any neck. For convenience we
decompose these into Fourier modes with wavevectors $q$ reciprocal to
the space variable $u$ along the transformed filament length:
\begin{equation}
	\begin{pmatrix} \edot(u,t)\\ a(u,t)	\\ Z(u,t)\\
	\end{pmatrix}
= 
	\begin{pmatrix}
		\edotbar_0(t)\\    a_0	\\		Z_0(t)\\
        \end{pmatrix}
+
	\sum_q\begin{pmatrix}
		\delta\edot(t)\\ \delta a(t)	\\	\delta Z(t)\\
	\end{pmatrix}_q \exp(iqu)
	. \label{eqn:matrixM}
\end{equation}
The area perturbations $\delta a(t)$ obey $\delta a(t)/a_0=\delta
A(t)/A_0(t)$ and accordingly measure the fractional variations in
cross sectional area along the filament's length, compared to the
length-averaged area, at any time $t$.  In this way, they measure the
degree of necking.

Expression (\ref{eqn:matrixM}) is then substituted into equations
(\ref{eqn:massT2}) to (\ref{eqn:visc2}). Expanding in successive
powers of the perturbation amplitude, and retaining only terms of
first order, then gives a set of linearised equations governing the
dynamics of the perturbations.

The linearised mass balance equation is
\be
\label{eqn:linMass}
\partials{\delta a_q}{t}=-\delta\edot_q.
\ee
The linearised force balance equation is
\be
\label{eqn:linForce2}
0=\sigmae \delta a_q + G\delta Z_q + \eta \delta\edot_q,
\ee
and the linearised viscoelastic constitutive dynamics
\be
\partials{\delta Z_q}{t}=\delta \edot_q f(Z_0)+C\delta Z_q.
\ee
In this equation the term
\be
\label{eqn:C}
C=\edot f'(Z_0)-\frac{1}{\tau}g'(Z_0),
\ee
in which a prime denotes differentiation with respect to a function's
own argument. 

The strain rate perturbations $\delta \edot_q$ are instantaneously
enslaved to the other variables by the condition of force balance in
creeping flow, Eqn.~\ref{eqn:linForce2}. Accordingly we can eliminate
these to arrive at the two-dimensional linear dynamical system:
\begin{equation}
\partial_t
        \begin{pmatrix}
		\delta a(t)	\\ \\	\delta Z(t)\\
	\end{pmatrix}_q 
= \tens{M}(t)
\cdot
        \begin{pmatrix}
		\delta a(t)	\\  \\ 	\delta Z(t)\\
	\end{pmatrix}_q, \label{eqn:2Dset}
\end{equation}
which is governed by the stability matrix
\be
\tens{M}(t)= 	\begin{pmatrix} \dfrac{\sigma_{\rm E0}}{\eta} & \dfrac{G}{\eta} \\
  & \\
  \dfrac{-f(Z_0)\sigma_{\rm E0}}{\eta}\;\;\;\ & -\dfrac{f(Z_0)G}{\eta} +C \\
  & \\
	\end{pmatrix}.
\label{eqn:MM}
\ee
We note that this matrix has inherited the (in general)
time-dependence of the base state $(\edotbar_0(t),a_0,Z_0(t))$, upon
which it depends.

The stability matrix $\tens{M}(t)$ does not however depend on the
wavevector $q$, and all Fourier modes $\exp(iqu)$ are predicted to
have the same dynamics.  We therefore expect the dominant mode in
practice to be determined by which is seeded most strongly initially.
We return to discuss this issue at the end of Sec.~\ref{sec:BCs} below.

\begin{figure*}
\centering			
\includegraphics[width=0.99\textwidth]{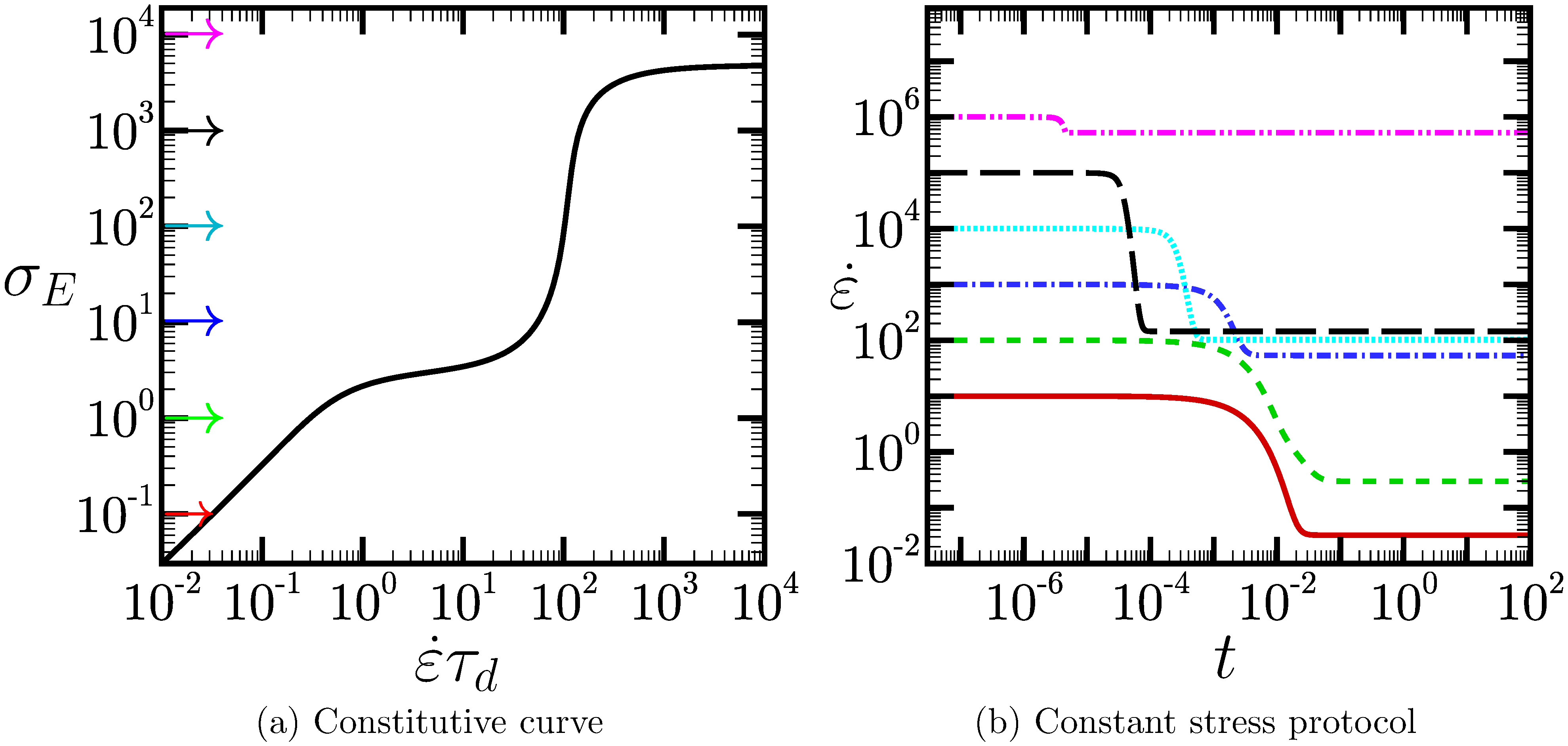}
\caption{\bf Finite-stretch Rolie-Poly model: (a) homogeneous stationary
  constitutive curve for extensional flow, (b) evolution of the
  strain-rate (of a homogeneous base state flow) to its value
  prescribed by the constitutive curve, following the switch-on of a
  constant stress in a previously undeformed sample. Arrows in (a)
  denote the imposed stress values, each colour-matched to its
  corresponding transient curve in (b), and with the initial strain
  rate values in (b) monotonic in the decreasing arrow locations in
  (a).}
\label{fig:transient}
\end{figure*}

\subsection{Nonlinear slender filament simulations}
\label{sec:simulations}

To study the dynamics of necking outwith the linear regime, once the
amplitude of the necking perturbations is no longer small, we evolve
the fully nonlinear slender filament equations~\ref{eqn:massT}
to~\ref{eqn:confT} numerically.  The fact that the length of the
filament remains fixed in the coextending frame in which those
equations are expressed, even as the sample stretches out in the
laboratory frame, obviates any need for re-meshing the numerical grid
over time.  Accordingly, we discretize the equations on a fixed mesh
and step the equations forward in time by means of an explicit Euler
algorithm for the spatially local terms and first order upwinding for
the convective ones~\cite{Press2007}. Details can be found in
Ref.~\cite{strainPaper}, along with a discussion of our approach to ensuring
convergence on the space and time-steps, which we also adopt here.

\subsection{Boundary conditions}
\label{sec:BCs}

In our linear stability calculations we assume periodic boundary
conditions between the two ends of the filament (implicitly taking the
filament to correspond to a torus being stretched).  In our nonlinear
slender filament simulations we use an approximate mimic of the
no-slip boundary condition between the fluid and the endplates, by
adopting an artificially divergent viscosity near each plate according
to Eqn.~VII.1 of Ref.~\cite{strainPaper}. As discussed in that reference, and at
the end of Sec.~\ref{sec:LSA} above, this automatically provides some
heterogeneity that seeds the formation of a neck mid-filament.
This in turn is likely to be set by the no-slip condition at the
rheometer plates, which constrains the area to remain constant at each
plate even as the sample is stretched out overall.  The overall effect
of this will be to initiate a single neck in the middle of the
filament.

\section{Constant stress protocol}
\label{sec:stress}

In this section we consider a filament of viscoelastic material that
is initially cylindrical and undeformed, with all internal stresses
well relaxed. It is then subject at some time $t=0$ to the switch-on
of a tensile stress $\sigmaei$, which is held constant thereafter.  In
response to this imposed stress, the most commonly measured bulk
rheological signal is the accumulated nominal Hencky strain $\ebar(t)$
as a function of the time $t$ since the stress was applied. This defines
the material's extensional creep curve, at the given stress.
Sometimes reported instead is its time-differential $\edotbar(t)$.
 
Our basic observation in this constant-stress protocol, which our
numerical results below will show, is that the sample's response
comprises two successive dynamical regimes that are (usually) quite
well separated. The first regime is one of fast dynamics at early
times after the imposition of the stress, during which the strain rate
$\edotbar(t)$ quickly proceeds to the value prescribed by the
material's stationary homogeneous constitutive curve at the given
imposed stress. After this the flow remains steady for the remainder
of this first regime.  During this first regime, the sample has little
or no time in which to develop a neck, and the dynamics can
accordingly be motivated by a simple calculation that considers only
the homogeneous (base state) dynamics.  We outline this in
Sec.~\ref{sec:fast}.

After that first fast regime, a second regime then ensues in which the
homogeneous flow state destabilises and the sample necks. We shall
derive analytical expressions for the rate at which the neck develops
in Sec.~\ref{sec:criteria}, and give numerical results supporting
these in Sec.~\ref{sec:stressNumerics}.

\subsection{Base state: fast evolution to constitutive curve}
\label{sec:fast}

As just noted, the dynamics at early times after the imposition of the
stress can be motivated by a calculation in which the flow is assumed
to remain homogeneous, to a good approximation, without necking. (The
nominal Hencky strain $\ebar$ therefore corresponds with the true
strain $\epsilon$ everyone along the filament, so we drop the overbar
for the rest of this subsection.) We now sketch the overall features
of this early-time response, first analytically in the context of our
simplified scalar toy model, then with our numerical results for the
full tensorial Rolie-Poly model.

The condition of force balance at constant imposed stress $\sigmae$ gives
\be
\label{eqn:homStress}
\sigmae=GZ + \eta\edot.
\ee
Within our scalar toy model, the viscoelastic variable $Z$
evolves according to
\be
\label{eqn:homDyn}
\dot{Z}=\edot f(Z)-\frac{1}{\tau} g(Z).
\ee
Instantaneously after the imposition of the stress the viscoelastic
contribution $GZ$ to the stress is zero and the load is all carried by
the solvent: a strain rate $\edot=\sigmaei/\eta$ therefore immediately
develops uniformly along the filament.
After this instantaneous initial response, the strain rate then
quickly evolves to its value on the constitutive curve at the given
stress. This can be motivated by differentiating Eqn.~\ref{eqn:homStress}
at constant $\sigmae=\sigmaei$ to get
\bea
\eddot &=& -\frac{G}{\eta}\dot{Z}\nonumber\\
&=& -\frac{G}{\eta}\left[\edot
  f(Z)-\frac{1}{\tau}g(Z)\right],\label{eqn:fast} 
\eea
where we have substituted $\dot{Z}$ from Eqn.~\ref{eqn:homDyn} in
moving from the first to the second line. In the second line, the
argument $Z$ of $f$ and $g$ is given by
$Z=\tfrac{1}{G}(\sigmae-\eta\edot)$ with constant $\sigmae=\sigmaei$.
In this way, Eqn.~\ref{eqn:fast} prescribes the evolution of the
strain rate $\edot$ on a fast timescale $\eta/G$ to its value on the
stationary homogeneous constitutive curve at the given imposed stress
$\sigmaei$.

We have motivated this early-time evolution here in the context of our
simple scalar model. However the same scenario arises in the full
tensorial constitutive models used throughout the rest of the paper.
Numerical results for the (homogeneous flow response of the) fully
tensorial Rolie-Poly model are shown as an example in
Fig.~\ref{fig:transient}b). For each imposed stress value, the
instantaneous appearance of a strain rate $\edot=\sigmaei/\eta$ at
time $t=0$ is clear, followed by fast evolution over a time interval
that scales as the short timescale $\eta/G$ to the strain rate
prescribed by the corresponding constitutive curve in
Fig.~\ref{fig:transient}a), for the given imposed stress.

\subsection{Linear instability: rate of necking}
\label{sec:criteria}

Following the fast evolution to a homogeneous flow state prescribed by
the stationary constitutive curve, as just described, a second
dynamical regime ensues in which that homogeneous flow state
destabilises and the sample starts to neck. We now perform a linear
stability analysis to determine the dynamics of the onset of this
necking process. To allow analytical progress, we do this within our
simplified scalar model. We intentionally leave the form of the
loading and relaxation functions $f(Z)$ and $g(Z)$ unspecified.  Our
aim in so doing is to motivate formulae for the rate of necking that
can be expressed in terms of characteristic features in the shape of
the underlying homogeneous constitutive curve, independently of any
specific constitutive choices for $f$ and $g$ (or their counterparts
in any fully tensorial model).

With the possibility of heterogeneous flow reinstated, the flow is now
governed by Eqns.~\ref{eqn:massT2} to~\ref{eqn:visc2}.  Within these
equations we consider a uniform time-independent base flow state
$\edot_0,a_0,Z_0$, as pertaining at the end of the early-time regime
described in Sec.~\ref{sec:fast}, with the strain rate $\edot_0$
prescribed by the underlying homogeneous constitutive for the given
imposed stress. To this initially homogeneous state we then add small
amplitude heterogeneous perturbations, as set out in
Eqn.~\ref{eqn:matrixM}, which are the precursor of a developing neck.
(Note that, although for the sake of pedagogy in this analytical
  section we describe the development of heterogeneity as starting
  only after the early fast evolution of Sec.~\ref{sec:fast} above, in
  practice these two regimes are only separated to good approximation.
  To account for any slight mixing of the two regimes, our numerics in
  Sec.~\ref{sec:stressNumerics} allow for heterogeneity right from the
  inception of the flow.)

Substituting Eqn.~\ref{eqn:matrixM} into Eqns.~\ref{eqn:massT2}
to~\ref{eqn:visc2}, expanding in powers of the amplitude of the
perturbations, and discarding any terms higher than those of first
order, gives the set of linearised equations~\ref{eqn:linMass}
to~\ref{eqn:C}, which govern the time evolution of the perturbations.
These can finally be expressed in matrix format as in
Eqns.~\ref{eqn:2Dset} and~\ref{eqn:MM}.  Because the base state
variables upon which the stability matrix $\tens{M}$ depends have
already attained stationary values on the underlying constitutive
curve by the end of the first fast regime, the matrix $\tens{M}$ that
we need to consider for this protocol is in fact time-independent.
Instability to necking will obtain in any regime where $\tens{M}$ has
at least one eigenvalue that is positive, in the sense of having
positive real part, with the associated rate of necking being
governing by the amplitude of that real part. Our aim now, therefore,
is to understand when any eigenvalue of $\tens{M}$ is positive, and
the way in which the rate of necking that it prescribes relates to any
quantities that could be measured experimentally.

The two eigenvalues of $\tens{M}$ follow as solutions of
\be
\omega^2-T\omega+\Delta =0,
\ee
where $T$ is the trace of $\tens{M}$ and $\Delta$ its determinant. 

The trace $T$ of $\tens{M}$ is given by
\be
\label{eqn:trace}
T=\frac{1}{\eta}(\sigmaeb-fG)=-\frac{1}{A(0)\eta}\EC.
\ee
(We shall return below to explain the meaning of the derivative $\EC$
in this expression.) The determinant $\Delta$ of $\tens{M}$ is given by
\be
\label{eqn:det}
\Delta=\frac{\sigmaeb C}{\eta} = -\frac{f}{\eta}\left[\frac{d\log\sigmaei}{d\edot}\right]^{-1}.
\ee
(Note that the solvent viscosity $\eta$ is small compared to the
viscoelastic viscosities, and we have ignored terms $O(1)$ compared to
those $O(G/\eta)$ in these expressions.) In each of
Eqns.~\ref{eqn:trace} and~\ref{eqn:det}, the second equality follows
from the first by a few lines of algebra, which we do not write down.

The first mode of necking instability arises in any regime where the
determinant $\Delta <0$, with a rate of necking per unit time that
scales as the inverse logarithmic derivative of the material's
underlying homogeneous constitutive curve:
\be
\label{eqn:ccm}
\omega\sim\left[\frac{d\log\sigmaei}{d\edot}\right]^{-1}\;\;\;\textbf{``Constitutive
  curve mode''}.  
\ee
This is a key result. It predicts that any material with a positively
sloping homogeneous extensional constitutive curve will be unstable to
necking \cite{Fielding2011,strainPaper}. Because the
vast majority of complex fluids and soft solids have such a curve, we
predict that essentially all materials will neck when subject to
extensional stretching. This prediction is indeed consistent with
ubiquitous reports of necking in the experimental literature.

The result in (\ref{eqn:ccm}) also tells us that a material will neck
relatively more quickly in any regime in which its homogeneous
constitutive curve of stress as a function of strain rate is
relatively flatter. This is intuitively understood as follows.
Consider an initially homogeneous filament. Then suppose that the
strain rate fluctuates upward slightly (compared to the averaged
imposed one) in some local region of the filament.  As a result of
this, the filament will thin a bit faster in that locality compared to
the filament as a whole. In order to maintain a uniform force along
the length of the filament, as required by the force balance
condition, a counterbalancing stress must be provided in the
developing neck to compensate for the thinned area. To generate this,
an even faster flow must develop in that slightly necked region. This
enhances the original fluctuation, giving positive feedback and
instability to necking.  The extent to which the material must indeed
flow faster to provide a counterbalancing stress in the developing
neck is determined by the inverse slope of its constitutive curve at
the relevant imposed stress, which therefore controls the rate of
necking according to Eqn.~\ref{eqn:ccm}.

In additional to the ``constitutive-curve'' mode just discussed, a
second mode predicts necking in any regime where the matrix $\tens{M}$
has trace $T>0$, and so where
\be
\label{eqn:EC}
\EC<0\;\;\;\textbf{``Elastic \considere mode''.}
\ee
The associated eigenvalue, which determines the rate of necking per
unit time associated with this mode, is $O(G/\eta)$, which is large
for the highly viscoelastic materials considered here.  This second
mode was also predicted in Ref.~\cite{strainPaper} as a route to necking
in the protocol of imposed Hencky strain rate. As discussed in
Ref.~\cite{strainPaper}, the derivative $\EC$ of the base state's tensile
force with respect to strain $\epsilon$ needs careful explanation.  It
is defined by evolving the system's state up to some strain $\epsilon$
with the full model dynamics, including loading by flow as encoded by
$f$ and relaxation as encoded by $g$. In the next increment of strain
$\epsilon \to \epsilon +\delta \epsilon$ over which the derivative is
taken the relaxation term $g(Z)$ is then suppressed, with only the
elastic loading dynamics implemented.  

Whether any means can be found of measuring this derivative
experimentally is an open question, which we do not address here.
However we do emphasise that the condition $\EC<0$ just discussed is
{\em not} the same as the condition $\partial_\epsilon F<0$. The
``elastic'' \considere criterion proposed here is therefore {\em not}
the same as the original \considere criterion for necking.  Indeed, in
any experiment performed at constant tensile stress $\sigmaei$ the
force $F=A\sigmaei$ decreases with increasing strain for all times
after the imposition of the load, because the stress is constant (by
definition) and the filament's area always thins with strain as
$A\sim\exp(-\ebar(t))$. Therefore, the original \considere criterion
entirely fails to predict necking in a constant stress protocol: it
must be replaced by the two new criteria offered here.

\subsection{Numerical results}
\label{sec:stressNumerics}

\subsubsection{Linear regime}

In the previous subsection we derived criteria for two different modes
of necking instability under conditions of constant imposed tensile
stress. To allow analytical progress, we did this in the context of
our simplified and highly generalised scalar constitutive model. In
this section, we shall present numerical results confirming the
validity of these criteria in six concrete choices of constitutive
model, and discuss in more detail the way in which the two modes of
necking manifest themselves in the various flow regimes of these
models.  The first five models (Oldroyd B, Giesekus, non-stretch
Rolie-Poly, finite-stretch Rolie-Poly and Pom-Pom) are fully tensorial
and popularly studied in the rheological literature.  The sixth is a
toy model of coil-stretch hysteresis, which we constructed so as to have
a non-monotonic constitutive curve, in order to demonstrate stability
against necking in the regime of negative constitutive slope
$\sigmae'(\edot)<0$.

\begin{figure*}
\centering	
\includegraphics[width=0.8\textwidth]{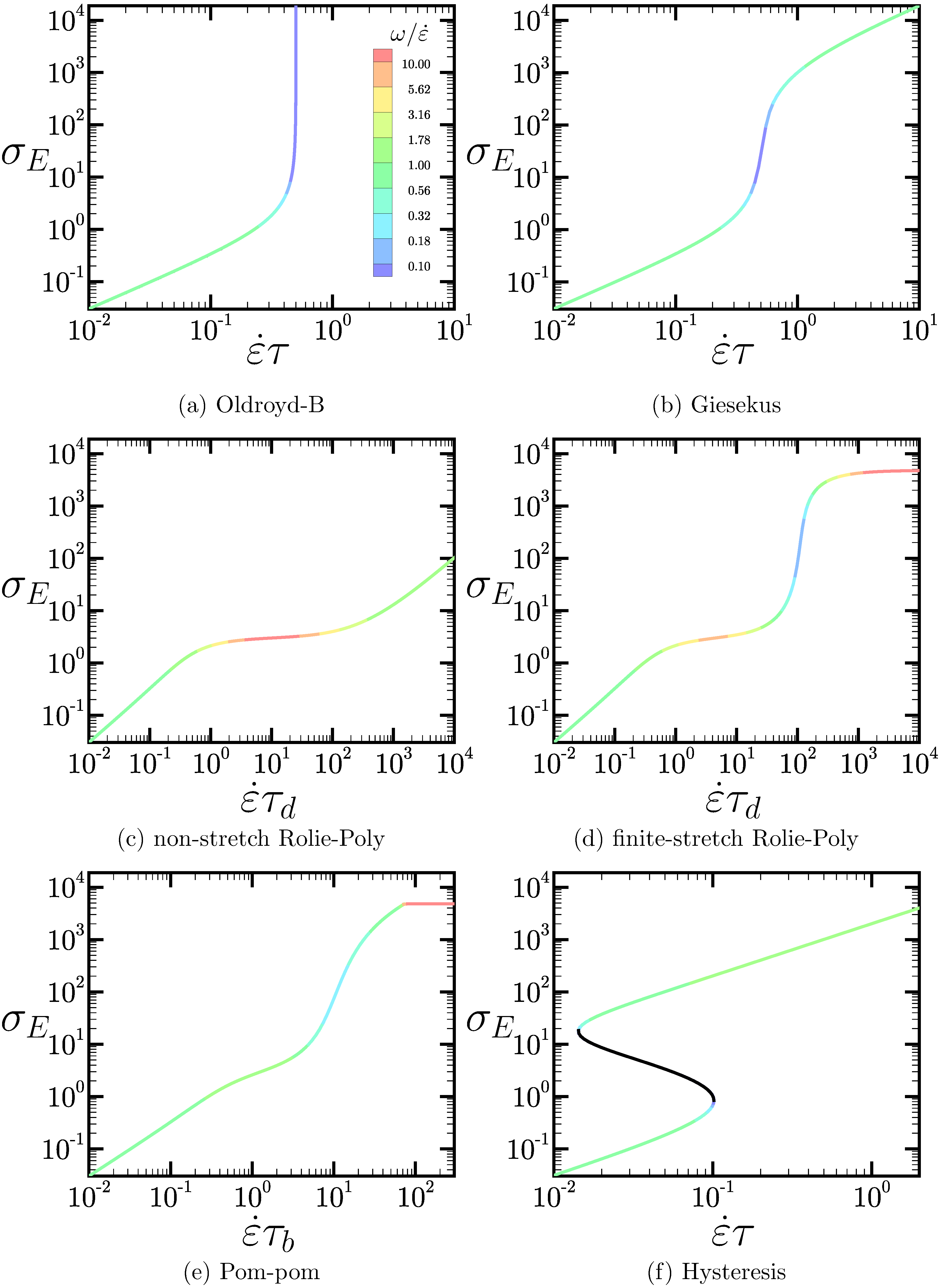}	
\caption{\bf Constitutive curves of the models to be studied. The colourscale in each case shows the value of $[d\log\sigmae/d\log\edot]^{-1}$, which sets the rate of necking per unit strain according to Eq.~\ref{eqn:ccms}. Any regime in which this quantity is less then or equal to zero, indicating stability against necking, is shown in black.}
\label{fig:constitutive}
\end{figure*}

We start in Fig.~\ref{fig:constitutive} by showing the underlying
homogeneous constitutive curve for each of the six models. As
discussed above, following the imposition of a constant stress
$\sigmaei$, the strain rate first evolves rapidly to its value as
prescribed by this underlying constitutive curve, with the sample
remaining (almost) uniform while that happens.  The constitutive curve
mode discussed above then predicts instability to necking in any
regime where that curve has positive slope. (As we shall find below,
the elastic \considere mode will also arise  in two of the models studied,
although in relatively restricted regimes of flow rate.)  Its onset
rate per unit time is given by Eqn.~\ref{eqn:ccm}.  Correspondingly,
its onset rate per unit strain is given by
\be
\label{eqn:ccms}
\frac{\omega}{\edot}\sim\left[\frac{d\log\sigmaei}{d\log\edot}\right]^{-1}\;\;\;\textbf{``Constitutive curve mode''},
\ee
that is, by the inverse of the derivative of the constitutive curve as
shown in a log-log representation.

This rate is indicated by the colourscale in each of the constitutive
curves in Fig.~\ref{fig:constitutive}: faster necking (as a function
of accumulated strain) is seen in regimes of flatter positive
constitutive slope. In the regime of negative constitutive slope of
the hysteresis model (panel f), stability against necking is
predicted.

\begin{figure*}
\centering	
\includegraphics[width=0.8\textwidth]{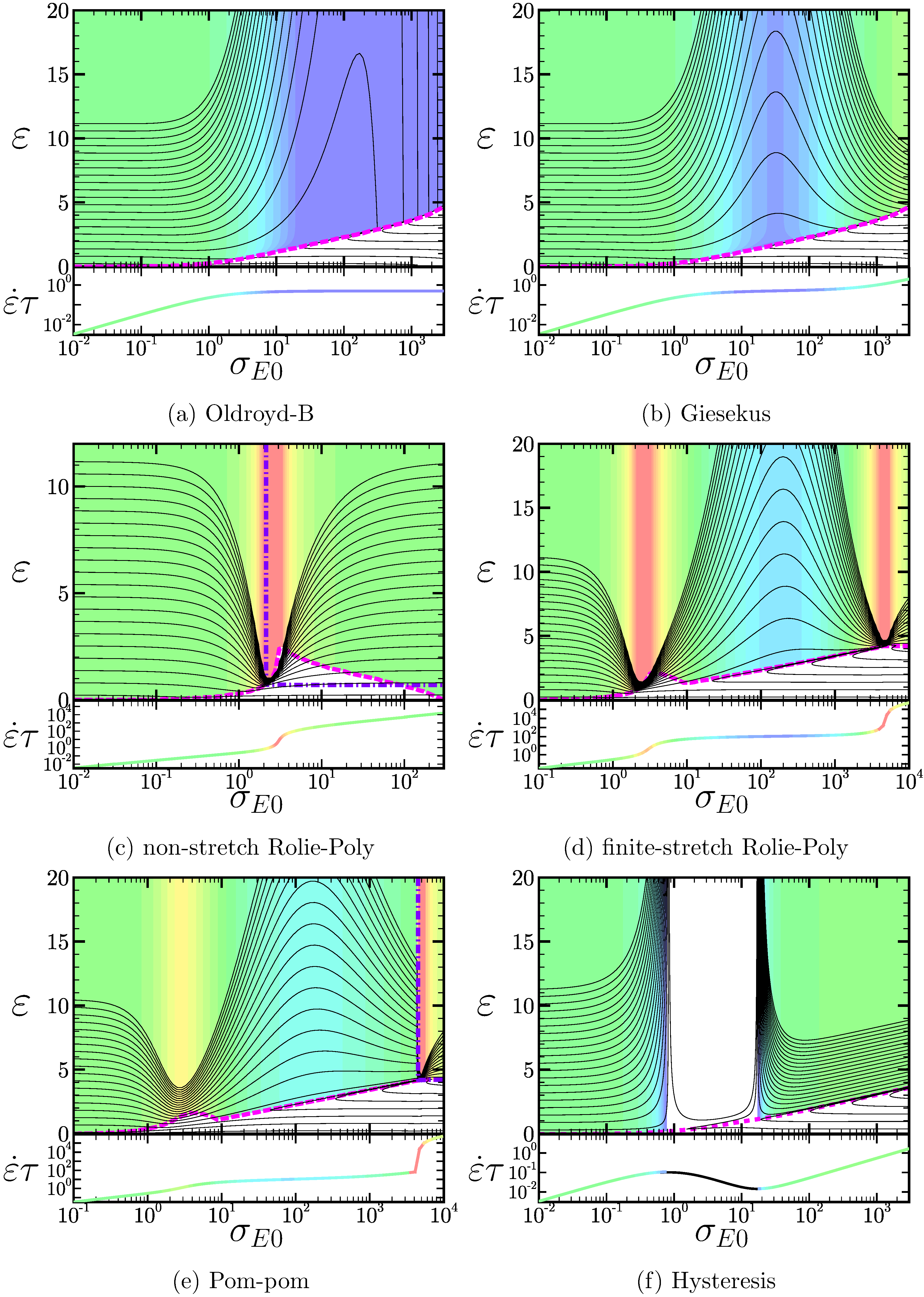}
\caption{\bf Numerical results for the linearised necking dynamics at
  constant imposed tensile stress within the six constitutive models.
  In each panel the lower subfigure shows the inverted constitutive
  curve, with the rate of necking (per unit strain) according to the
  constitutive curve mode shown on the same colourscale as in
  Fig.~\ref{fig:constitutive}. The upper subfigure in each panel
  explores the necking dynamics in more detail. Here the thin black
  lines show contours of constant area perturbations $\delta a /
  \delta a_0 = 10^{n/4}$, with $n = 1 \cdots 20$ in curves from bottom
  to top, representing the growing degree of necking at increasing
  strain $\epsilon$ upwards in any filament stretching experiment at
  fixed imposed stress $\sigmaei$.  The magenta dashed lines show the
  strain at which the underlying base state attains a state of
  stationary flow on the underlying homogeneous constitutive curve, to
  within $1\%$.  The purple dot-dashed line shows the strain at which
  the elastic \considere mode becomes unstable. The colours denote the rate of necking, with the same colourscale as in \ref{fig:constitutive}. Regions of stability against necking are shown in white.
  }
\label{fig:stress}
\end{figure*}

With these general remarks in mind, we now explore in detail the
necking dynamics of these six models under conditions of constant
imposed tensile stress. Each panel (a)-(f) of Fig.~\ref{fig:stress}
corresponds to its counterpart constitutive curve panel in
Fig.~\ref{fig:constitutive}, and comprises two separate subgraphs.
The lower subgraph in each case essentially reproduces the
corresponding constitutive curve of Fig.~\ref{fig:constitutive}, but
with the axes inverted so as to have the stress on the abscissa, as
the imposed quantity in this protocol. At any imposed stress, the
expected rate of necking as a function of accumulated strain is again
shown by the colourscale according to Eqn.~\ref{eqn:ccms}.

The upper subgraph in each panel of Fig.~\ref{fig:stress} shows the
necking dynamics in detail. The data are presented in the plane
$(\ebar,\sigmae)$ of accumulated strain and imposed stress, and should
be interpreted as follows. Any vertical cut up this plane corresponds
to single experiment in which the imposed tensile stress $\sigmae$ is
held fixed and the accumulated strain $\ebar$ increases up the plane
as the filament stretches out under the influence of this applied
load.  At any imposed stress value $\sigmae$, the magenta dashed line
shows the strain at which the strain rate attains its value on the
stationary underlying constitutive curve, to within $1\%$, following
the fast early-time dynamics described in Sec.~\ref{sec:fast}.

The solid black lines show contours of constant area perturbation
$\delta a(t)$, with each successive contour crossing corresponding to
an increase by a factor $10^{1/4}$ in the degree of necking $\delta
a(t)$. The $nth$ contour thus represents a degree of necking $\delta a
/ \delta a_0 = 10^{n/4}$, where $\delta a_0$ is the small initial
seeding at the start of the run.  The more densely clustered the
contour lines vertically at any fixed $\sigmae$, therefore, the faster
necking occurs in an experiment at that imposed stress.  In each panel
of Fig.~\ref{fig:stress} we have shown only the first $20$ contour
lines, assuming that the sample will have failed altogether by this
time.  An indication of the dependence of the strain at which the
sample will finally fail on the imposed stress is given by focusing on
one representative contour.

Consistent with the claim made in Sec.~\ref{sec:fast} above, only a
small amount of necking occurs (few contour lines are crossed) during
the fast early-time regime in which the strain rate rapidly evolves to
the stationary underlying constitutive curve. Beyond that early
regime, the neck indeed develops (contour lines are crossed) at a rate
consistent with the scaling prediction of (\ref{eqn:ccms}), set by the
inverse of the slope of the underlying constitutive curve in its
log-log representation. We now discuss in more detail the way this
basic observation manifests itself in the different flow regimes of
each model.

In the Oldroyd B model, the constitutive curve has a constant initial
slope $\sigmae'(\edot)=3$ in the Newtonian regime at low strain rates.
See Fig.~\ref{fig:constitutive}a. The associated rate of necking per
unit strain is likewise modest and constant as a function of strain
rate, as seen in Fig.~\ref{fig:stress}a).  For higher imposed stresses
the constitutive curve is a much steeper function of strain rate, due
to the well-known extensional catastrophe of this model, which we
recall predicts a divergent stress for strain rates $\edot>1/2\tau$.
This greatly stabilises the filament against necking, as seen by the
near vertical contour lines in Fig.~\ref{fig:stress}a for larger
imposed stresses.

In the Giesekus model the divergent constitutive curve of the Oldroyd
B model is avoided and the stress remains finite at all strain rates.
Nonetheless, a vestige of the Oldroyd B catastrophe is seen in the
steeply sloping section at strain rates $\edot\approx 1/2$ in
Fig.~\ref{fig:constitutive}b. Accordingly, the model predicts a regime
of relatively slow necking for imposed stresses $10^1<\sigmae<10^3$ in
that regime of steep constitutive slope, and so of relatively flat
strain rate versus stress. This is indeed seen in that stress range
$10^1<\sigmae<10^3$ in the contour map of Fig.~\ref{fig:stress}b).
Faster necking is seen for imposed stresses on either side of this
window, where the constitutive curve of Fig.~\ref{fig:constitutive}b
is Newtonian at low flow rates, or quasi-Newtonian at high flow rates.

The constitutive curve of the non-stretch Rolie-Poly model
(Fig.~\ref{fig:constitutive}c) has a Newtonian regime at low strain
rates, in which chain orientation progressively increases with
increasing applied strain rate. Once the strain rate
$\edot=O(1/\taud)$, however, the orientation saturates and the stress
becomes a much flatter function of strain rate, rising again only at
much higher strain rates when the solvent contribution $3\eta\edot$
becomes important. For imposed stresses $\sigmae\approx 3.0$
corresponding to the flat region where $\edot=O(1/\taud)$, therefore,
we expect very rapid necking via the constitutive curve mode. This is
confirmed in Fig.~\ref{fig:stress}c).

In fact for imposed stresses $\sigmae\gtrsim 2.0$ the elastic
\considere mode is also active, as indicated by the purple dot-dashed
line.  According to Eqn.~\ref{eqn:EC}, this predicts a rate of necking
$O(G/\eta)$ per unit time, which is very fast. (Recall that the
solvent viscosity $\eta$ is small compared to the zero shear viscosity
of the viscoelastic components.)  However it is important to note that
the corresponding rate of necking per unit strain is $O(G/\eta\edot)$.
Now the constitutive curve of this model, in this regime where chain
orientation has saturated, is dominated by the solvent contribution.
Accordingly the associated strain rate $\edot=\sigmae/\eta$.
Combining these gives a rate of necking $O(G/\sigmae)=O(1)$ per unit
strain, which remains relatively gentle. This is indeed confirmed by
our numerical results to the right of Fig.~\ref{fig:stress}c): because
both the rate of necking and the rate of straining are fast,
$O(G/\eta)$, the rate of necking per unit strain remains $O(1)$.

The Rolie-Poly model with chain-stretch now included is explored in
Figs.~\ref{fig:constitutive}d) and~\ref{fig:stress}d).  For strain
rates $\edot<1/\taur$, its dynamics essentially match that of its
non-stretch counterpart discussed in the previous paragraph,
consistent with the fact that negligible chain stretch develops for
imposed flow rates lower than the rate of chain stretch relaxation.
Once the strain rate $\edot=O(1/\taur)$, however, significant chain
stretch develops and the constitutive curve stress rises rapidly as a
function of strain rate. For imposed stresses corresponding to this
regime, much slower necking is predicted: note the blue colourscale in
Fig.~\ref{fig:constitutive}c).  This is indeed observed, via the
widely spaced contour lines for imposed stresses in the range
$10^1-10^3$ in Fig.~\ref{fig:stress}d).  At higher stresses still the
chain stretch saturates, the constitutive curve is much flatter, and
the necking is much faster.

The dynamics of the finite-stretch Rolie-Poly model can therefore be
categorised into four distinct regimes, as follows. For low imposed
stresses, corresponding to strain rates in the regime $\edot<1/\taud$
in Fig.~\ref{fig:constitutive}d), the model shows essentially
Newtonian response: the constitutive curve has a constant slope
$\sigmae'(\edot)=3$, and the rate of necking is likewise modest and
independent of stress. For imposed stresses $\sigmae\approx 3.0$ the
chain orientation has saturated and the constitutive curve is a
relatively flat function of strain rate, giving very fast necking.
For imposed stresses in the range $10^1-10^3$ the constitutive curve
has large slope due to the development of chain stretch in that regime,
and the rate of necking is accordingly much lower. Finally at high
stresses the chain stretch saturates, the constitutive curve is much
flatter, and necking occurs very quickly.  We note that the elastic
\considere mode, which was present in the non-stretch Rolie-Poly
model, is absent in the stretching version of the model.

The Pom-Pom model is explored in Figs.~\ref{fig:constitutive}e)
and~\ref{fig:stress}e). Comparing Figs.~\ref{fig:constitutive}d)
and~\ref{fig:constitutive}e), we see that the constitutive curve of
the Pom-Pom model also shows four regimes, which resemble those of the
finite-stretch Rolie-Poly model. Indeed in close analogy with the
dynamics of the Rolie-Poly model, these regimes are associated with a
progressive increase then saturation in backbone orientation, for
strain rates $\edot=O(1/\tau_b)$ and their counterpart imposed
stresses, followed by an increase then saturation in backbone stretch,
for strain rates $\edot=O(1/\tau_s)$ and their counterpart imposed
stresses.

The Pom-Pom model accordingly shows very similar necking dynamics to
those of the Rolie-Poly model, as can be seen by comparing
Figs.~\ref{fig:stress}d) and~\ref{fig:stress}e).  However at the
highest imposed stresses there is an importance difference between the
two models: in the Pom-Pom model the backbone stretch is assumed to
have an infinitely sharp cutoff, compared with the more gentle
saturation in the chain stretch of the Rolie-Poly model. This gives an
essentially flat regime in the constitutive curve at the right hand
side of Fig.~\ref{fig:constitutive}e), which manifests itself in
Fig.~\ref{fig:stress}e) as very violent necking via the elastic
\considere mode. Whether this cutoff is physically realistic remains
an open question, which we discussed in the context of necking under
conditions of constant imposed Hencky strain rate in Ref.~\cite{strainPaper}.

Finally in Fig.~\ref{fig:stress}f) we explore the necking
  dynamics of our toy model of coil-stretch hysteresis, for which the
  underlying constitutive curve is non-monotonic
  (Fig.~\ref{fig:constitutive}f).  The important point to note here is
  that filament is stable against necking for imposed stresses in the
  regime of negative slope in that constitutive curve. Whether a
  different mode of extensional failure would take over in practice
  remains an open question.

\begin{figure}
\centering	
\includegraphics[width=0.45\textwidth]{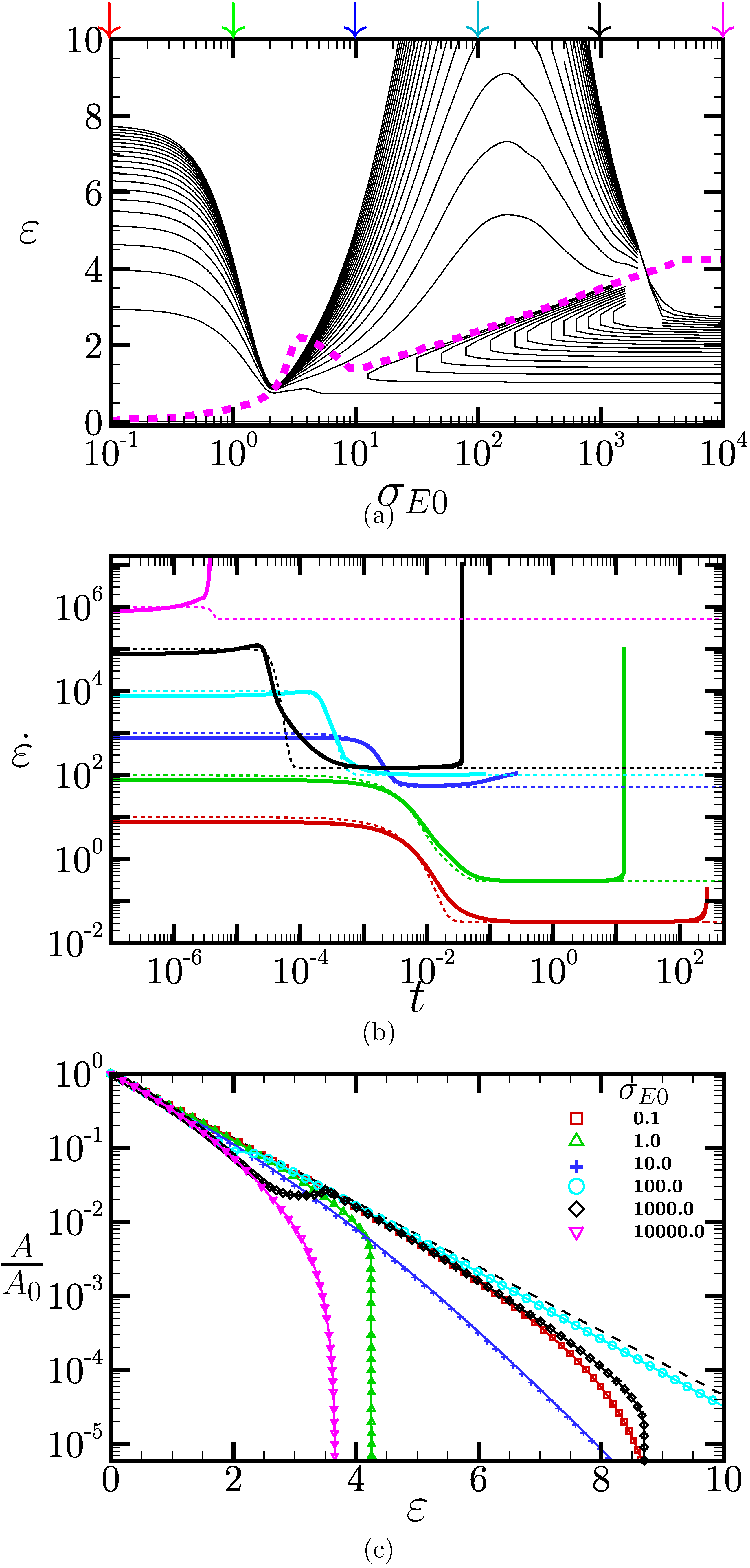}	
\caption{\bf Necking dynamics in nonlinear slender filament simulations of
  the Rolie-Poly model of linear entangled polymers under conditions
  of constant imposed tensile stress. (a) Thin black lines show
  contours of constant of necking heterogeneity, $\Lambda$, with the n$^{th}$ 
  contour having $\Lambda = 4^{n/20}$. Also shown by the
  magenta dashed line is the strain at which the sample attains a
  state of flow on the homogeneous constitutive curve to within $1\%$,
  which occurs before significant necking develops. The sharp corners in some 
  earlier contours are an artefact of a breakdown in our interpolation. 
  For the six imposed strain rates indicated by arrows in (a), the evolution of
  the nominal strain rate as a function of time since the inception of
  the flow is reported in (b) for both the nonlinear simulation (solid
  lines) and for a calculation in which the filament is artificially
  assumed to remain uniform (dashed lines).  Counterpart results for
  the filament's cross sectional area are shown in (c): assuming
  homogeneous flow (dashed black line) and at the filament's midpoint
  in a calculation that allows for necking (symbols and solid lines).
  }
\label{fig:nonlinear}
\end{figure}

\subsubsection{Nonlinear dynamics}

So far we have discussed necking dynamics in the linear regime, in
which the amplitude of the growing necking perturbations still remains
small.  To study the necking dynamics outwith the linear regime, once
the amplitude of the heterogeneous perturbations becomes
non-negligible, we simulated the full nonlinear slender filament
equations.

Results for the finite-stretch Rolie-Poly model are shown in
Fig.~\ref{fig:nonlinear}. Panel a) shows the equivalent, for these
nonlinear calculations, of the linear stability results discussed
above in Fig.~\ref{fig:stress}d). As usual, a vertical cut up this
plane of strain $\epsilon$ and stress $\sigmaei$ represents a single
experiment performed at a given imposed tensile stress $\sigmaei$. The
experiment starts with an initially unstretched filament at
$\epsilon=0$, then the accumulated strain $\epsilon$ increases up the
plot as the filament progressively stretches out.

The thin black lines then show contours of constant $\Lambda(t)\equiv
A_{\rm hom}(t)/A_{\rm mid}(t)$, where $A_{\rm hom}(t)$ is the filament
area calculated at any time $t$ by supposing the filament were
stretching in a purely uniform way, and $A_{\rm mid}(t)$ is the actual
cross sectional area at the filament's midpoint.  In this way,
$\Lambda=1$ corresponds to a uniform filament, and $\Lambda$
progressively increases as the filament necks.  The first contour has
$\Lambda=1$, and each successive contour as $\epsilon$ increases at
fixed $\sigmaei$ indicates an increase in $\Lambda$ by a factor
$4^{1/20}$. The $20th$ contour, which is the final shown, therefore
represents $\Lambda=4$.  (Although this ratio of areas is relatively
modest we note that the sample is close to necking by this time,
because the global area has become very small.)  Accordingly we show
only the first $20$ contour lines, and simply assume that the sample
will fail altogether by some final pinch-off event before this contour
is attained. We note, however, that our slender filament calculation
is not capable of capturing the dynamics of the final pinchoff,
because it assumes variations along the filament to be gentle on the
scale of the filament radius. It also neglects any physics on the
microscopic lengthscales associated with the surface tension of the
interface, which would surely affect the final stages of failure.

Comparing Fig.~\ref{fig:nonlinear}a) with Fig.~\ref{fig:stress}d), we
see that our simplified linear calculation in fact already performed
rather well in predicting the onset of necking in the full nonlinear
calculations. Any differences in the quantitative detail between the
two plots can be explained by the fact that the material functions at
the filament's midpoint in the nonlinear simulations differ slightly
from those in the base state of the linear calculation.

The necking dynamics at the six imposed stress values denoted by
arrows in Fig.~\ref{fig:nonlinear}a) are presented in further detail
in panels b) and c). Panel b) shows the evolution of the nominal
(length-averaged) strain rate as a function of time since the
inception of the flow for the nonlinear simulation (solid lines), and the
strain rate predicted by a calculation in which the filament is
artificially assumed to remain uniform (dashed lines).  In each case
except that of the highest imposed stress, the strain rate attains its
value prescribed by the homogeneous constitutive curve with only
minimal necking occurring transiently en route to that curve. A second
flow regime then ensues in which this state of homogeneous flow on the
constitutive curve destabilises and the strain rate increases as the
filament necks.  Counterpart results for the cross sectional area at
the filament's midpoint are shown in (c). This follows the exponential
prediction of the homogeneous calculation until necking occurs, when
the signal plunges below the homogeneous predictions. The least
  necking arises for the imposed stress value $\sigmaei=100.0$, which
  corresponds in the constitutive curve Fig.~\ref{fig:constitutive}d)
  to a strain rate of order $1/\taur$, in the regime where chain
  stretch is rapidly developing with increasing strain rate. Indeed,
  at this stress value the flow remains on the underlying constitutive
  curve without any appreciable necking effects up to the final strain
  $\epsilon=10.0$ considered.  The most dramatic necking in
  Fig.~\ref{fig:nonlinear}c) arises for an imposed stress
  $\sigmaei=1.0$, in the regime of saturated orientiation in the
  constitutive curve of Fig.~\ref{fig:constitutive}d).
Fig. \ref{fig:nonlinear}(b) demonstrates that for larger imposed tensile 
stresses there is a smaller time-window over which the strain rate accords
 with its value as predicted by the stationary homogeneous constitutive curve, 
 before necking causes it to deviate significantly.  This has been observed 
 experimentally: see, for example, Fig. 9 of Ref. \cite{Munstedt2013}.

\section{Constant force protocol}
\label{sec:force}

\begin{figure*}
\centering	
\includegraphics[width=0.99\textwidth]{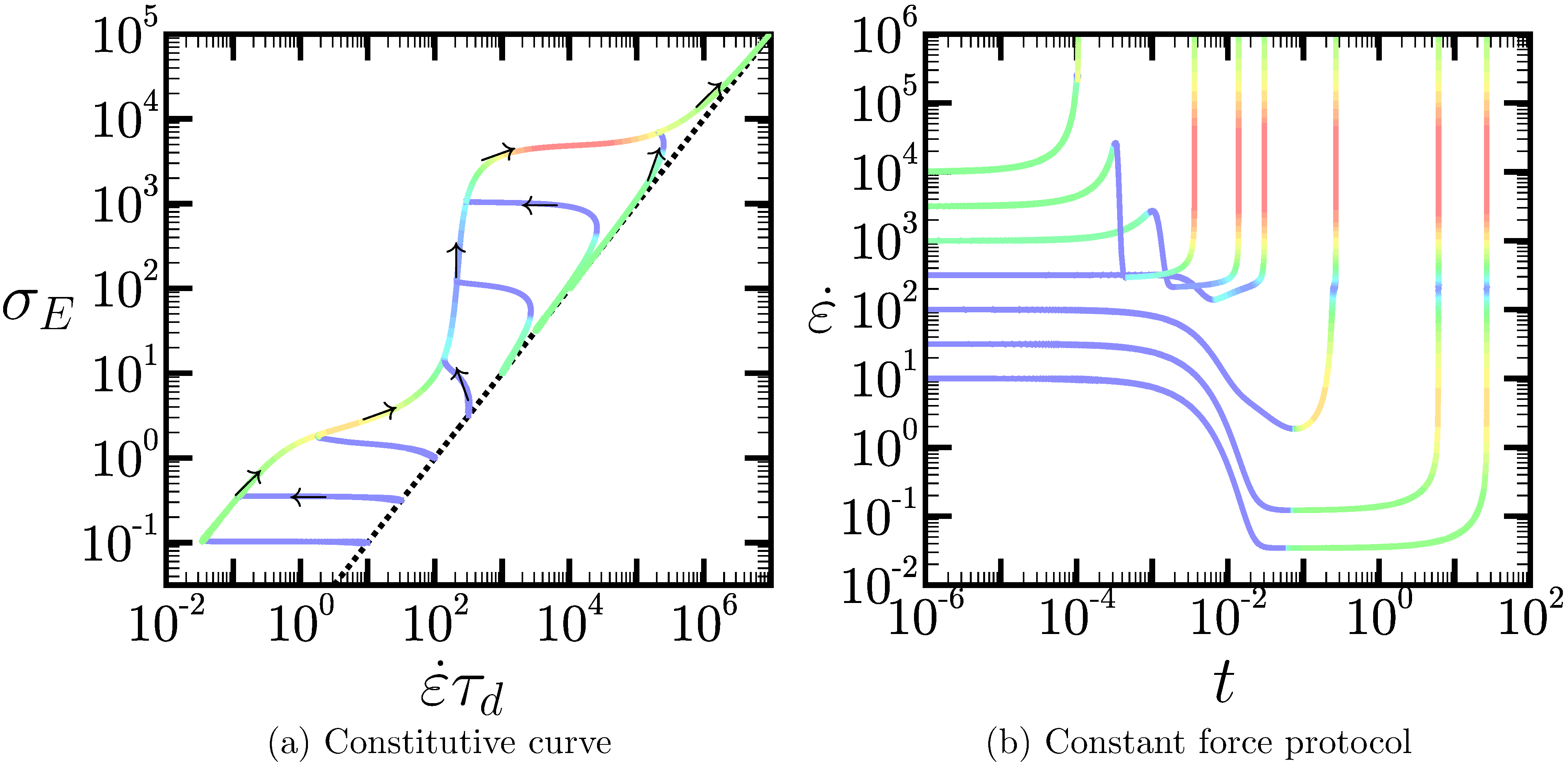}	
\caption{\bf Finite-stretch Rolie-Poly model. (a) Viscoelastic
  constitutive curve (leftmost limiting curve line) and solvent
  constitutive curve (dotted line). The trajectory of the base state
  flow is also shown, for several different imposed forces, in this
  flow-curve representation of stress as a function of strain rate (b)
  Evolution of the strain-rate with time. Imposed constant forces are $10^{n/2}$ 
  with $n = -2 \cdots 4$ and the colourscale is the same as in figure 
  \ref{fig:constitutive} showing the value of $[d\log\sigmae/d\log\edot]^{-1}$.}
\label{fig:transientForce}
\end{figure*}

In this section we consider a filament of viscoelastic material that
is initially cylindrical and undeformed, with all internal stresses
well relaxed. At some time $t=0$ it is then subject to the switch-on
of a tensile force $\Fi$, which is held constant thereafter.

For the protocol of constant imposed stress in the previous section,
we showed that (to good approximation) the sample attains a state of
stationary flow specified by the homogeneous extensional constitutive
curve, at the given imposed stress, before starting to neck. That
enabled us to consider the emergence of a necked state out of an
initially homogeneous `base state' that was stationary in time.

In important contrast, under conditions of constant force we shall
find that the development of the neck happens in tandem with a
continuous time evolution in the underlying homogeneous base state out
of which the necked state emerges. In what follows, therefore, our
strategy will be first (in Sec.~\ref{sec:base}) to discuss the
time-evolution of that homogeneous underlying base state, and then (in
Sec.~\ref{sec:neck}) to discuss the way in which a neck develops as
that base state evolves.  As usual our analytical discussion will
focus for the sake of simplicity and generality on the toy scalar
constitutive model, with numerical calculations confirming the same
scenario in the fully tensorial models.

\subsection{Base state evolution}
\label{sec:base}

Within the toy scalar model, the homogeneous base state obeys
Eqns.~\ref{eqn:baseMass} to~\ref{eqn:baseZ}. Repeating these here for
convenience, we have the condition of mass balance
\be
\label{eqn:massHere}
\dot{A}_0(t)=-\edot_0 A_0,
\ee
while the tensile force, which is time-independent in this protocol, obeys
\be
\label{eqn:forceHere}
F_0=\Fi=A_0\sigma_0=A_0(GZ_0+\eta\edot_0),
\ee
with $\sigma_0$ the tensile stress. The viscoelastic variable obeys
\be
\label{eqn:viscHere}
\dot{Z}_0(t)=\edot_0f(Z_0)-\frac{1}{\tau}g(Z_0).
\ee

Immediately following the imposition of a force $\Fi$, the stress
$\sigma_0(0)=\Fi/A_0(0)$, where $A_0(0)$ is the filament's initial
cross sectional area.  The viscoelastic contribution to the stress is
initially zero, and all the load is carried by the solvent.
Accordingly a uniform initial strain rate $\edot_0=\Fi/A_0(0)\eta$
instantaneously establishes along the filament. This is seen in the
initial values of the strain-rate signal versus time in
Fig.~\ref{fig:transientForce}b), which shows our numerical results for
the base-state dynamics of the Rolie-Poly model. In the flow-curve
representation of Fig.~\ref{fig:transientForce}a), therefore, the base
state's trajectory starts (for any given imposed force) on the solvent
constitutive branch $\sigmae=\eta\edot$, which is shown as a dotted
black line.

In contrast to the protocol of constant imposed stress, in this
imposed-force protocol the stress must progressively increase in time
in order to maintain a constant force as the filament stretches out
and its cross sectional area thins.  (Recall that the tensile force is
the product of the tensile stress and the cross sectional area.) This
can be seen by differentiating Eqn.~\ref{eqn:forceHere} to get
\be
\dot{F}_0(t)=0=\dot{A}_0\sigma_0+A\dot{\sigma}_0,
\ee
which rearranges to give
\be
\label{eqn:stressIncrease}
\frac{\dot{\sigma}_0}{\sigma_0}=-\frac{\dot{A}_0}{A_0}=\edot_0.
\ee
The second equality here follows from the mass balance
condition~\ref{eqn:massHere}. Hence, the stress increases in time with
a fractional growth rate given at any time $t$ by the strain rate
$\edot_0(t)$.

If the flow were to remain dominated by the Newtonian solvent, with no
load transfer to the viscoelastic component, this strain rate
$\edot_0(t)$ would remain trivially given by
$\edot_0(t)=\sigma_0(t)/\eta$ for all times, and the fractional rate
of stress increase would accordingly likewise equal
$\sigma_0(t)/\eta$.  In the flow-curve representation of
Fig.~\ref{fig:transientForce}a), the base-state's trajectory would
then simply sweep up the solvent constitutive branch shown by the
black dotted line. At any time $t$, however, there is a competing
tendency of the flow to evolve away from the solvent constitutive
branch to the composite constitutive curve of the combined
viscoelastic and solvent components (which we hereafter call simply
the viscoelastic constitutive curve), at a rate $G/\eta$.  Recall
Eqn.~\ref{eqn:fast} and the discussion immediately following it.

For typical initial stress values $\sigma_0\lesssim G$, the second of
these processes (of evolution to the viscoelastic constitutive curve)
is faster than the first (of time-increase of the stress up the
solvent branch). As a result, for these low imposed force values the
system's trajectory in the flow-curve representation of
Fig.~\ref{fig:transientForce}a) quickly evolves away from the solvent
constitutive branch and leftwards to the viscoelastic constitutive
curve, before having chance to sweep any way up the solvent branch.
This progression from the Newtonian to viscoelastic constitutive curve
is associated with a fast sudden decrease in the signal of strain rate
as a function of time $\edot_0(t)$ at a time $t\approx 10^{-2}$ for
the three lowest imposed force values in
Fig.~\ref{fig:transientForce}b).  Thereafter the base state flow
remains specified by the viscoelastic constitutive curve, exploring it
in an upward direction as the stress increases according to
Eqn.~\ref{eqn:stressIncrease}, and the strain rate likewise rapidly
increasing in Fig.~\ref{fig:transientForce}b). (In fact this
discussion holds for models with a finite extensional viscosity.  The
Oldroyd B model exhibits different, but pathological behaviour on
account of its extensional catastrophe.)

For initial stress values $\sigma_0\gtrsim G$, in contrast, the
time-increase of the stress up the solvent constitutive branch is
initially faster than the evolution away from that branch. Accordingly
for these larger imposed force values the base state trajectory in the
flow-curve representation of Fig.~\ref{fig:transientForce}a) first
explores the solvent constitutive curve in an upward direction, and
only later makes its transit leftwards to the viscoelastic
constitutive curve, with the associated sudden decrease in strain
rate.  Once the viscoelastic constitutive curve has been attained, the
flow again explores it with increasing stress and strain rate
according to Eqn.~\ref{eqn:stressIncrease}.

In any such regime where the base state flow sweeps up one of these
underlying constitutive branches (whether initially that of the
solvent or finally the full viscoelastic constitutive curve), the time
rate of change of the stress is given by
\be
\dot{\sigma}_0=\sigma_0'(\edot_0)\frac{d\edot_0}{dt},
\ee
where the function $\sigma_0(\edot_0)$ specifies the constitutive
curve in question, and $\sigma_0'(\edot_0)$ its slope.  Substituting
this into Eqn.~\ref{eqn:stressIncrease}, and rearranging, gives an
expression for the time rate of change of the strain rate:
\be
\label{eqn:strainRateChange}
\frac{1}{\edot_0}\frac{d\edot_0}{dt}=\frac{\sigma_0}{\sigma_0'(\edot_0)}.
\ee
In any regime where the constitutive curve in question is
approximately Newtonian, $\sigma_0\propto\edot_0$, this can easily be
shown to lead to a blow-up of the strain rate in a finite time set by
the inverse initial strain rate:
\be
\edot_0(t)=\frac{\edot_0(0)}{1-\edot_0(0)t}.
\ee
This is indeed evident in the signals of strain rate as a function of
time in Fig.~\ref{fig:transientForce}b). 

Although the blow-up happens quickly in terms of time, however, the
filament nonetheless develops large strains during this blow-up. In
what follows, indeed, our primary focus will be on elucidating the
rate at which a neck develops as a function of the accumulating strain
(while also recognising that the corresponding rate of necking as a
function of elapsed time will in many cases be very quick indeed).

\subsection{Rate of necking}
\label{sec:neck}

So far, we have discussed how a base state of homogeneous flow evolves
in time under conditions of constant imposed force. In this section,
we consider how a necked state develops out of that initially
homogeneous time-evolving base state flow.  At the level of linearised
dynamics the heterogeneous perturbations $\delta\edot_q(t),\delta
a_q(t),\delta Z_q(t)$ that are the precursor of any neck evolve
according to Eqns.~\ref{eqn:linMass} to~\ref{eqn:C}, which we repeat
here for convenience. The linearised mass balance condition gives
\be
\label{eqn:onea}
\partials{\delta a_q}{t}=-\delta\edot_q.
\ee
The linearised force balance condition gives
\be
\label{eqn:twoa}
0=\sigmae \delta a_q + G\delta Z_q + \eta \delta\edot_q,
\ee
and the linearised viscoelastic dynamics give
\be
\label{eqn:threea}
\partials{\delta Z_q}{t}=\delta \edot_q f(Z_0)+C\delta Z_q,
\ee
in which
\be
C=\edot f'(Z_0)-\frac{1}{\tau}g'(Z_0).
\ee

Our aim in what follows is to relate the dynamics of these necking
perturbations ($\delta\edot_q(t),\delta a_q(t),\delta Z_q(t)$) to the
time-evolution of the underlying base state
$\edot_0(t),A_0(t),Z_0(t)$.  Doing so will allow us to make
predictions for the rate at which a necked state develops, in terms of
characteristic signatures in the rheological signals of the base
state. It is also important to note that, before any significant
necking occurs, the rheological signals of the base state match the
bulk rheological signals measured experimentally. Therefore, our
predictions for the onset of a necked state are indeed made in terms
of characteristic signatures in the experimentally measured
rheological quantities.

\begin{figure*}
\centering	
\includegraphics[width=0.89\textwidth]{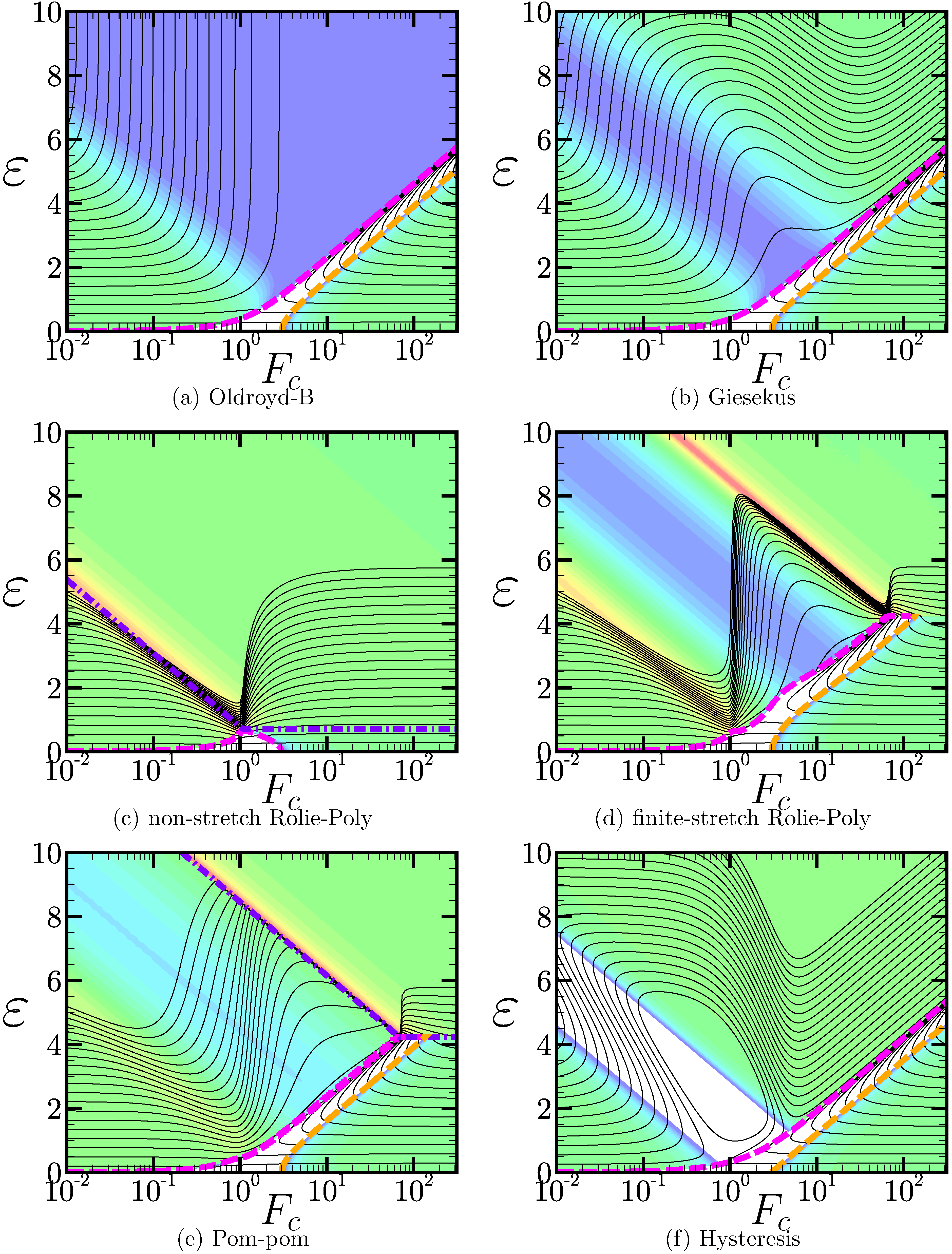}
\caption{\bf Numerical results for the linearised necking dynamics at
  constant imposed tensile force within the six constitutive models.
  In each panel the thin black lines show contours of constant area
  perturbations $\delta a / \delta a_0 = 10^{n/8}$, with $n = 1 \cdots
  20$ in curves from bottom to top,
  representing the growing degree of necking at increasing strain
  $\epsilon$ upwards in any filament stretching experiment at fixed
  imposed force $\Fi$.  The orange dashed line shows the strain at which
  the underlying base state first departs the solvent constitutive
  branch by more than $1\%$. The magenta dashed lines show the strain
  at which the underlying base state attains the underlying
  homogeneous viscoelastic constitutive curve, to within $1\%$.  The
  purple dot-dashed line shows the strain at which
  the elastic \considere mode becomes unstable. The colours show the
  rate of necking for any $(\ebar,\Fi)$, using the same colourscale as in figure 
  \ref{fig:constitutive}. Regions of stability against necking are shown in white.}
\label{fig:force}
\end{figure*}

The time-evolution of the base state is specified by
Eqns.~\ref{eqn:massHere} to~\ref{eqn:viscHere}. Differentiating these
with respect to time (with the mass balance condition first
pre-divided by $A_0(t)$) gives
\be
\label{eqn:oneb}
\frac{d}{dt}\left(\frac{\dot{A}_0}{A_0}\right)=-\frac{d\edot_0}{dt},
\ee
\be
\label{eqn:twob}
\frac{dF_0}{dt}=0=\sigma_0\frac{dA_0}{dt}+A_0\left(G\frac{dZ_0}{dt}+\eta\frac{d\edot_0}{dt}\right),
\ee
\be
\label{eqn:threeb}
\frac{d\dot{Z}_0}{dt}=\frac{d\edot_0}{dt}f(Z_0)+C\frac{dZ_0}{dt}.
\ee

Comparing Eqns.~\ref{eqn:onea},~\ref{eqn:twoa} and~\ref{eqn:threea}
with Eqns.~\ref{eqn:oneb},~\ref{eqn:twob} and~\ref{eqn:threeb}, we
note that the heterogeneous perturbations $(\delta\edot_q(t),\delta
a_q(t),\delta Z_q(t))$ to the homogeneous base state obey the same
dynamical equations as the base state quantities
$(\ddot\epsilon_0,\dot{A}_0/A_0,\dot{Z}_0)$.
Therefore~\footnote{Technically, the equality~\ref{eqn:forceNeck} only
  holds if the initial conditions of $(\delta\edot_q(t),\delta
  a_q(t),\delta Z_q(t))$ are the same as those for
  $(\ddot\epsilon_0,\dot{A}_0/A_0,\dot{Z}_0)$. This will not in
  general be the case. However, after an initial transient associated
  with any discrepancy in the initial conditions, we find that
  Eqn.~\ref{eqn:forceNeck} indeed holds well in all the constitutive
  models that we have studied.}, this means that

\be
\label{eqn:forceNeck}
\frac{1}{\delta a_q}\frac{d\delta a_q}{dt}=\frac{d}{dt}\left(\frac{\dot{A}_0}{A_0}\right)/\frac{\dot{A}_0}{A_0}=\frac{1}{\edot_0}\frac{d\edot_0}{dt},
\ee
in which the final equality follow from the mass balance
condition~\ref{eqn:massHere}.  This is an important result: it
predicts that any initially small heterogeneous perturbations along the
filament's length will grow towards a necked state in any regime where
the bulk strain rate signal $\edot_0(t)$ -- {\it i.e.}, the
time-differential of the creep curve -- increases in time. A converse
tendency to return towards heterogeneous flow is predicted in any
regime where the strain rate decreases in time.

Looking back at Fig.~\ref{fig:transientForce}, we recall that the
strain rate indeed increases in time in any regime where the sample
sweeps up any branch of a (positively sloping) constitutive curve.
The associated rate of necking per unit strain can be calculated by
substituting Eqn.~\ref{eqn:strainRateChange} for the rate of change of
strain rate into Eqn.~\ref{eqn:forceNeck} to get
\be
\label{eqn:forceNeck1}
\frac{1}{\edot_0}\frac{1}{\delta a_q}\frac{d\delta
  a_q}{dt}=\frac{1}{\edot_0^2}\frac{d\edot_0}{dt}=\frac{\sigma_0}{\edot_0}\frac{d\edot_0}{d\sigma_0}=\left(\frac{d\log
    \sigma_0}{d\log\edot_0}\right)^{-1} \ee

At any instant during the evolution of the (base state) flow up one of
the underlying constitutive branches, therefore, the rate of necking
per unit strain is given by the inverse slope of that constitutive
curve on a log-log plot. Indeed this mirrors the result of
Eqn.~\ref{eqn:ccms} for the rate of necking (per unit strain) under
conditions of constant imposed stress.  The important difference in
this case of constant imposed force is that the base state doesn't
remain at a fixed point on the constitutive curve, but rather moves
upward along that curve as the filament's area thins and the stress
accordingly increases to maintain constant force.

In motivating the result in Eqn.~\ref{eqn:forceNeck1}, we
  implicitly assumed (via our use of Eqn.~\ref{eqn:strainRateChange})
  that the flow state is instantaneously prescribed by the underlying
  constitutive curve at any value of the upwardly evolving stress.
  While this is true to good approximation, the flow in fact must obey
  the dynamical viscoelastic constitutive equation.  In making that
  assumption, therefore, we effectively neglected one mode of the
  system's dynamics.  Correctly accounting for the viscoelastic
  constitutive dynamics gives another mode of necking instability, the
  rate of which is set by the inverse of the same quantity $\EC$ as in
  Eqn.~\ref{eqn:EC} above: we identify this as the manifestation of
  the elastic \considere mode in the constant force protocol.

\subsection{Numerical results}

With the predictions of the previous section in mind, we now present
our linear stability results for necking in the constant force
protocol.  These are shown in Fig.~\ref{fig:force}, each panel (a)-(f)
of which corresponds to a counterpart constitutive curve panel in
Fig.~\ref{fig:constitutive}.  The data are presented in the plane
$(\ebar,\Fi)$ of accumulated strain and imposed force, which should be
interpreted as follows. Any vertical cut up this plane corresponds to
single experiment in which the imposed tensile force $\Fi$ is held
fixed and the accumulated strain $\ebar$ increases up the plane as the
filament stretches out under the influence of this load.

The orange and magenta dashed lines in Fig.~\ref{fig:force} reflect our
discussion of the trajectory of the base state flow in
Fig.~\ref{fig:transientForce}a) above. In particular, at any imposed
force $\Fi$ the orange dashed line shows the strain at which the base
state flow first departs by more than $1\%$ from the solvent
constitutive branch. The magenta dashed line shows the strain at which
it attains the stationary underlying viscoelastic constitutive curve,
to within $1\%$. (These lines merge towards the right hand side of the
figure for the finite-stretch Rolie-Poly and Pom-pom models: at very
high stresses in these models the constitutive curve is dominated by
the solvent contribution.)

The solid black lines show contours of constant area perturbation
$\delta a(t)$. Each successive contour corresponds to an increase in
the degree of necking $\delta a(t)$ by a factor $10^{1/8}$. The $nth$
contour thus represents a degree of necking $\delta a / \delta a_0 =
10^{n/8}$, where $\delta a_0$ is the small initial seeding at the
start of the run.  The more densely clustered the contour lines
vertically at any fixed $\Fi$, therefore, the faster necking occurs in
an experiment at that imposed force.

Focusing first for definiteness on our results for the Giesekus model
in panel b), we see that for any imposed force $\Fi\lesssim 1.0$ the
base state flow quickly transits to the viscoelastic constitutive
curve as the strain increases vertically up the panel at that fixed
$\Fi$, without first exploring the Newtonian branch.  Indeed, the
scale of strain over which this transit happens is barely discernible.
After attaining the viscoelastic constitutive curve the flow
progressively sweeps up it as the filament stretches out with
accumulating strain $\ebar(t)$, thins in area as $\exp(-\ebar(t))$,
and the stress correspondingly increases to maintain a constant force.
At any accumulated strain $\ebar$, for the given imposed $\Fi$, the
colourscale in Fig.~\ref{fig:force}b) shows the inverse logarithmic
slope of the constitutive curve at the stress that has been attained
by that strain, matching the colourscale in
Fig.~\ref{fig:constitutive}b). Recall from Eqn.~\ref{eqn:forceNeck1}
that this sets the rate at which the neck will be developing at any
time $t$, or correspondingly accumulated strain $\ebar(t)$, during the
stretching process. For example, for a representative imposed force
$\Fi=0.1$ in Fig.~\ref{fig:force}b), the flow quickly attains the
viscoelastic constitutive curve of Fig.~\ref{fig:constitutive}b) at a
stress $\sigmae\approx 0.1$. It then transits up the slow-flow
Newtonian part of of this viscoelastic constitutive curve, with an
$O(1)$ rate of necking indicated in green, then transits the steeply
sloping part it, with a slower rate of necking indicated in blue, then
finally transients the Newtonian fast-flow part of it, with a return
to the $O(1)$ rate of necking indicated in green.

In contrast, for a representative imposed force $\Fi=100.0$ in
Fig.~\ref{fig:force}b), the base state flow initially evolves up the
solvent constitutive branch $\sigmae=\eta\edot$ (not shown in
Fig.~\ref{fig:constitutive}b), with an $O(1)$ rate of necking
indicated in green. Between a strain of roughly $4$ and $5$ it then
makes a rapid transit to the Newtonian fast-flow part of the
viscoelastic constitutive curve in Fig.~\ref{fig:constitutive}b),
during which the degree of necking actually decays slightly.  Finally,
it transients up that fast-flow part of the viscoelastic constitutive
curve, again with an $O(1)$ rate of necking indicated in green.

Although we have focused the discussion here on the Giesekus model for
the sake of a definite example, a closely analogous explanation
underpins essentially all the results in Fig.~\ref{fig:force}.
To summarise: under conditions of a constant imposed force the flow
typically sweeps up the underlying viscoelastic constitutive curve of
the material. At any given time, the rate of necking is given by the
inverse of the slope of that constitutive curve, shown in a log-log
plot, at the stress that has been attained by that time. For low
imposed forces the flow first sweeps up the constitutive curve of the
solvent contribution, before transiting to the viscoelastic curve.

One further feature appears in the results for the non-stretch
  Rolie-Pole and Pom-Pom models in Figs.~\ref{fig:force}c) and e)
  respectively: here the elastic \considere mode also operates, and is
  marked as a purple dot-dashed line on the figures.

\section{Conclusions}
\label{sec:conclusions}

By means of linear stability analysis and nonlinear simulations
performed at the level of a slender filament approximation, we have
studied the onset of necking during the stretching of an initially
cylindrical filament of complex fluid or soft solid, separately under
conditions of constant imposed tensile stress and constant imposed
tensile force. Our results pertain to highly viscoelastic filaments of
large enough radius that bulk stresses dominate surface effects, with
surface tension neglected accordingly.

Under conditions of constant imposed tensile stress, the flow first
quickly attains a state with the strain rate prescribed by the
underlying homogeneous stationary extensional constitutive curve of
the fluid in question, at the given imposed stress value.  During this
early time regime, the filament remains homogeneous to good
approximation, without any significant necking.  A second regime then
ensues in which that initially homogeneous flow destabilises to the
formation of a neck. This instability can occur via one of two modes,
the first of which arises widely across all the constitutive models
that we have studied, while the second is rarer in comparison.

The first mode of instability has a characteristic rate of necking per
accumulated Hencky strain unit set by the inverse of the slope of the
underlying stationary homogeneous constitutive curve, on a log-log
plot, at the given imposed stress. This is an important result:
essentially all materials of which we are aware have a positively
sloping extensional constitutive curve, and should therefore be
unstable to necking under conditions of a constant imposed tensile
stress. This prediction is indeed consistent with ubiquitous reports
of necking in the literature.

A possible rare exception to this ``constitutive curve'' mode of
necking is however predicted in a material that has a non-monotonic
extensional constitutive curve, for imposed stresses in the regime
where that curve has negative slope, such as could arise in polymer
solutions displaying coil-stretch hysteresis~\cite{Somani2010,Schroeder2003,DeGennes1974}. It
would be interesting to study the implications of the analysis offered
here in that case. It would also be interesting to consider the
implications of the presence of a yield stress in a material for
necking (or its possible absence) at imposed stresses below yield,
where the constitutive curve is essentially vertical. 
Necking in constant strain rate and constant velocity protocols in a yield stress
fluid were discussed previously in \cite{Hoyle2015}.

Under these conditions of a constant imposed stress a second, more
dramatic mode of necking instability can also arise. This is
essentially the direct counterpart of the elastic \considere mode
predicted earlier by ourselves in the context of filament stretching
at constant imposed Hencky strain rate~\cite{strainPaper}. However it
arises relatively rarely: of the six constitutive models studied in
this work, it only occurs in only two of them, and in a relatively
restricted regime of strain rate in each case. In particular, it
arises in models of polymeric flow in which polymer chain stretch is
dramatically inhibited: in the Rolie-Poly model of linear entangled
polymers with chain stretch disallowed, and in the Pom-Pom model of
branched entangled polymers with a hard cutoff in the permitted degree
of chain stretch.

We have also studied the onset of necking under conditions of constant
imposed tensile force. In this protocol, in contrast to the case of
constant tensile stress, the stress progressively increases over time
in order to maintain a constant tensile force as the filament
stretches out and gets thinner.  (Recall that the force is the product
of the stress and the filament's cross sectional area.)  Typically,
the flow simply sweeps progressively up one of the fluid's underlying
constitutive branches (whether that of the solvent, or the full
viscoelastic constitutive curve).  At any time during this progression
the rate of necking is set by the inverse of the slope of that
constitutive curve (on a log-log plot), at the stress that has been
attained by that time, in close analogy with the case of a constant
imposed stress.  Accordingly, a neck is predicted to develop in any
regime in which the measured strain rate signal increases in time.
During progression up a Newtonian or quasi-Newtonian part of a
constitutive curve, a finite-time divergence is predicted in the
signal of strain rate (and associated rate of necking) as a function
of time. The elastic \considere mode also operates in the non-stretch
Rolie-Poly and Pom-Pom models in some regimes.

Throughout this manuscript we have restricted ourselves to regimes
relevant to our slender filament approximation, in which the
wavelength of any heterogeneities that develop along the length of the
filament are long compared with the filament radius. The manner in
which these `fluid-like' necking instabilities cross over to more
dramatic solid-like fracturing at very high imposed loads, where the
sample sharply rips across its cross section, remains an interesting
open issue. 

{\it Acknowledgements} -- The research leading to these results has
received funding from the European Research Council under the European
Union's Seventh Framework Programme (FP7/2007-2013) / ERC grant
agreement number 279365.

\end{document}